\documentclass[a4paper,11pt]{article}
\ifdefined\XeTeXversion\else\pdfoutput=1\fi
\usepackage{jheppub}
\usepackage[T1]{fontenc}
\usepackage{amssymb}
\usepackage{graphicx}

\numberwithin{equation}{section}
\newcommand{\aQQ}{\alpha_{QQ}}
\newcommand{\aQj}{\alpha_{Qj}}
\newcommand{\ajj}{\alpha_{jj}}
\newcommand{\aQ}{\alpha_{Q}}
\newcommand{\aj}{\alpha_{j}}
\newcommand{\NQQ}{N_{QQ}}
\newcommand{\Njj}{N_{jj}}
\newcommand{\Cset}{\mathcal C}
\newcommand{\GG}{\mathcal G}
\newcommand{\Dom}{\mathcal D}
\newcommand{\AF}{\mathcal A_5}
\newcommand{\Rch}{\mathcal R}
\newcommand{\Hor}{\mathcal H}
\newcommand{\fp}{\mathfrak p}
\newcommand{\dist}{\operatorname{dist}}
\newcommand{\Zp}{\mathbb Z_{\ge0}}
\newcommand{\hj}{\hat\jmath}
\newcommand{\tauh}{\hat\tau}
\newcommand{\bmin}{\beta_{\min}}
\newcommand{\Qmin}{Q_{\min}}
\newcommand{\jmin}{j_{\min}}
\newcommand{\jst}{j_{\rm st}}

\title{Near-horizon BF bounds for the $t^+$ KK tower on extremal AdS$_5$ black holes}

\author[a]{Xiao-Long Liu,}
\author[a]{Zheng Miao,}
\author[b,c]{Yuezhang Tang}

\affiliation[a]{Department of Physics, Lehigh University,\\16 Memorial Drive East, Bethlehem, PA 18015, USA}
\affiliation[b]{Interdisciplinary Center for Theoretical Study, University of Science and Technology of China,\\Hefei, Anhui 230026, China}
\affiliation[c]{Peng Huanwu Center for Fundamental Theory,\\Hefei, Anhui 230026, China}

\emailAdd{xila25@lehigh.edu}
\emailAdd{zhm323@lehigh.edu}
\emailAdd{yuezhangtang@ustc.edu.cn}

\preprint{USTC-ICTS/PCFT-26-77}

\abstract{We determine the charged Breitenlohner--Freedman (BF) bound for every horizon harmonic of the type IIB $t^+$ Kaluza--Klein tower on extremal charged AdS$_5$ black holes with equal rotations. Starting from the covariant scalar equations of maximal gauged supergravity and exceptional field theory, we express the BF discriminant as a quadratic form in charge and spin. For charge aligned with the rotation, this form changes signature on the supersymmetric (BPS) curve, where every mode satisfies the bound. Above the curve, every black hole has a violating mode, but no finite cutoff in level or spin detects violation throughout this region. Below the curve, charge and spin discreteness permits satisfaction of the bound within a thin strip adjoining the BPS curve. At large radius and fixed rotation, the minimum occurs at level four with spin proportional to the radius, and the bound is violated even where all horizon singlets satisfy it. These results concern one tower at exact extremality in two-derivative supergravity; a dynamical instability requires a separate radial analysis.}


\begin{document}

\maketitle
\raggedbottom

\section{Introduction}\label{sec:intro}

The near-horizon geometry of an extremal black hole provides a tractable setting in which to study the competition between the energy of a perturbation, its charge and its angular momentum. Separation into horizon harmonics reduces the scalar equation to that of a charged field on AdS$_2$, with an effective mass and charge determined by the geometry~\mbox{\cite{Bardeen:1999px,Kunduri:2007vf,Kunduri:2013gce}}. The Breitenlohner--Freedman (BF) bound and its charged extension characterize when the AdS$_2$ scaling exponents are real~\mbox{\cite{Breitenlohner:1982bm,Breitenlohner:1982jf,PiolineTroost:2005AdS2}}. Scalar condensation and rotational instabilities provide physical motivations for this test~\mbox{\cite{Durkee:2010ea,Cardoso:2004nk,Gubser:2008px,Denef:2009tp}}.

We consider the Chong--Cveti\v{c}--L\"u--Pope (CCLP) black holes of five-dimensional minimal gauged supergravity~\mbox{\cite{Cvetic:2004hs,Chong:2005hr}}. These backgrounds lift to type IIB supergravity with AdS$_5\times S^5$ asymptotics~\mbox{\cite{Cvetic:2000nc}}. For equal rotations, the horizon is a squashed three-sphere, whose Hopf circle is twisted over AdS$_2$ in the near-horizon geometry. A horizon harmonic has SU(2) spin $j$ and fiber momentum $k$, with $|k|\le j$. Angular variation contributes positive gradient energy, but the fiber momentum also changes the effective AdS$_2$ charge through the rotational twist. The lowest BF discriminant therefore need not belong to the constant harmonic $j=k=0$.

The internal Kaluza--Klein level and the horizon harmonic describe different excitations. In the $t^+$ tower, the field at internal level $n\ge0$ has charge $Q=n+2$ and mass squared $M^2=n^2-4$, in units of the AdS radius~\mbox{\cite{Kim:1985ez,Gunaydin:1984fk,Eloy:2026jzg}}. Each such field admits all horizon harmonics. Ezroura and Larsen studied the level-zero supergravity fluctuations and their near-horizon couplings~\mbox{\cite{Ezroura:2024xba}}. Exceptional field theory provides a systematic description of the higher Kaluza--Klein levels~\mbox{\cite{Hohm:2013vpa,Baguet:2015sma,Malek:2019eaz,Malek:2020yue}}. Within this framework, Eloy et al. obtained the covariant equations for the $t^+$ tower, including its vanishing Pauli coupling, and analyzed the U(2)-invariant horizon singlets~\mbox{\cite{Eloy:2026jzg}}. Their analysis leaves regions near the BPS curve and at large radius where every singlet satisfies the bound. Whether those regions survive the remaining horizon harmonics is the question addressed here.

Starting from these covariant equations, we determine the BF bound for every horizon harmonic throughout the tower. Completing the harmonic test reveals which singlet-passing regions persist and identifies candidate perturbations for a radial analysis. We first consider charge aligned with the rotation, on the domain $\Dom$ with CCLP charge parameter $q_C\ge0$, rotation $0<a<1$ and horizon radius $r_+>0$.

The supersymmetric Gutowski--Reall solutions lie on $r_+^2=a(a+2)$, where the horizon angular velocity $\Omega_H$ and electric potential $\Phi_H$ both equal one~\mbox{\cite{Gutowski:2004ez,Gutowski:2004yv,Kunduri:2006ek}}. Above this curve in the $(r_+,a)$ plane, $\Omega_H>1$; below it, $\Omega_H<1$ and $\Phi_H>1$~\mbox{\cite{Ezroura:2024xba}}. We use $\beta$ for the charged BF discriminant and call a black hole BF-passing when every mode of the tower obeys $\beta\ge0$.

Minimization over the fiber momentum selects $k=-j$ on $\Dom$, leaving an exact quadratic polynomial in charge and spin. Its quadratic part is indefinite above the BPS curve and positive definite below it; on the curve, every mode satisfies the bound. Above the curve, $\beta$ is unbounded below over the tower. Near a BPS rotation without a saturating mode, detecting violation requires both the internal level and the horizon spin to increase without bound. Below the curve, the region $\beta<0$ in the real charge--spin plane is an ellipse, and a black hole passes precisely when this ellipse misses the allowed lattice. The passing set is confined to a thin strip adjoining the BPS curve. We determine fixed-level boundaries, bound the width of the passing region and characterize its near-BPS behavior.

\begin{figure}[tbp]
\centering
\includegraphics[width=\textwidth]{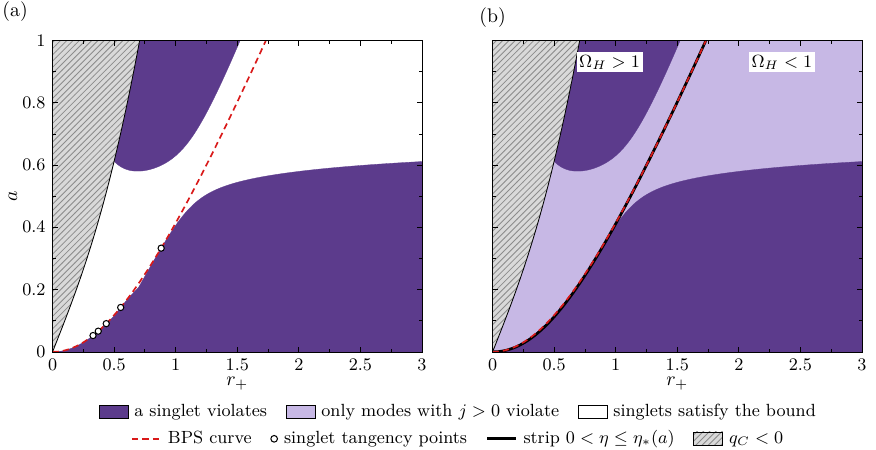}
\caption{The near-horizon BF bound in the $(r_+,a)$ plane, with horizon radius horizontal and rotation vertical. The red dashed BPS curve separates $\Omega_H>1$ (above and left) from $\Omega_H<1$, $\Phi_H>1$ (below and right). (a) Dark purple marks violation by a U(2) singlet; white means that every singlet satisfies the bound. Circles mark singlet saturation points with $n\le4$. (b) Light purple adds violation by nonzero-spin harmonics. The black curve is the outer enclosure $\eta=\eta_*(a)$ of section~\ref{sec:width}, where $\eta=\sqrt{1+2a^2+2r_+^2}-1-2a$ measures displacement from the BPS curve. The whole-tower passing set lies between these curves, within a strip too narrow to resolve here. Singlet shading follows minimization over integer charge, while the additional shading uses the above-BPS result and the below-BPS outer bound. Gray hatching marks the continuation to $q_C<0$, where charge is anti-aligned with rotation (section~\ref{sec:antialigned}).}
\label{fig:overview}
\end{figure}

At large radius and fixed rotation, the minimizing internal level remains $n=4$, while the spin grows proportionally to $r_+$. The minimum approaches $-3$, recovering the zero-momentum BF discriminant of the extremal planar Reissner--Nordstr\"om brane in this tower~\mbox{\cite{Gubser:2009qm}}. Angular harmonics therefore also remove the singlet BF-passing region at large radius. For the two extremal families with charge opposite to the rotation, treated in section~\ref{sec:antialigned}, the bound is violated throughout their domains. Figure~\ref{fig:overview} summarizes the change from the singlet analysis.

The analysis concerns one tower at exact extremality in two-derivative supergravity. Establishing a growing mode requires the full radial equation with its asymptotic AdS boundary condition; passing the BF test likewise does not establish radial stability. Limits involving unbounded quantum numbers require additional control when the corresponding momenta approach the string scale.

Section~\ref{sec:setup} introduces the backgrounds, scalar equation and BF criterion. Section~\ref{sec:form} derives the charge--spin quadratic form, and section~\ref{sec:susy} analyzes the fixed-level boundaries near the BPS curve. Section~\ref{sec:wholetower} treats the complete tower, followed by the large-radius limit and anti-aligned families in section~\ref{sec:large}. Section~\ref{sec:disc} discusses the physical implications. Appendix~\ref{app:A} derives the near-horizon equation and verifies decoupling and Pauli cancellation. Appendices~\ref{app:B} and~\ref{app:D} give the fixed-level and whole-tower bounds, respectively.

\section{Black hole backgrounds and the near-horizon BF criterion}\label{sec:setup}

\subsection{Extremal equal-rotation CCLP black holes and the BPS curve}\label{sec:branch}

We consider the CCLP solutions of minimal gauged supergravity with equal rotation parameters $a=b$ and unit AdS radius~\mbox{\cite{Chong:2005hr}}.\footnote{\label{fn:conv}We use mostly plus signature and set the gauge coupling to $g=1$. We denote the nonrotating CCLP coordinates by $(t_C,r,\theta_C,\phi_C,\psi_C)$, with $0\le\theta_C\le\pi/2$ and $\phi_C,\psi_C$ of period $2\pi$. The $t^+$ tower couples to $\AF=A_{\rm CCLP}/\sqrt3$, with horizon potential $\Phi_H$.} An extremal black hole is labeled by the rotation parameter $a\in(0,1)$ and the horizon radius $r_+>0$. It is convenient to replace $r_+$ by $s=\sqrt{1+2a^2+2r_+^2}>1$. In terms of $a$ and $s$, the CCLP charge and mass parameters are
\begin{equation}
\label{eq:qm}
q_C=\tfrac12(1+s)(s^2-s-2a^2),\qquad m_C=\tfrac18(s^2-1)(3s^2+1-4a^2).
\end{equation}

At fixed $(a,r_+)$, extremality fixes $m_C$ and leaves two roots for the charge parameter $q_C$. One is $q_C=s\,r_+^2-a^2$, which agrees with \eqref{eq:qm}; the other is $q_C=-a^2-s\,r_+^2$, obtained from \eqref{eq:qm} by $s\to-s$. All the resulting solutions are regular black holes (appendix~\ref{app:nhdata}). We restrict to the domain $\Dom=\{0<a<1,\ r_+>0,\ q_C\ge0\}$. Its boundary $q_C=0$ is the extremal Kerr--AdS$_5$ black hole with equal rotations, and $\Dom$ contains all supersymmetric black holes in this equal-rotation family. Because of the Chern--Simons term, $\AF\to-\AF$ is a symmetry of minimal gauged supergravity only together with a reversal of orientation, which maps $a=b$ to $a=-b$. The black holes with $q_C<0$ therefore carry charge anti-aligned with the rotation.

Supersymmetry requires $q_C=m_C/(1+2a)$. Using \eqref{eq:qm}, this condition becomes
\begin{equation}
\label{eq:doubleroot}
(1+2a)\,q_C-m_C=-\tfrac18(1+s)\,\eta^2\,(3s-1+4a),\qquad \eta\equiv s-1-2a .
\end{equation}
Supersymmetric black holes therefore lie on the BPS curve $\eta=0$, i.e.\ $r_+^2=a(a+2)$, and correspond to a double root of this condition. We use $\eta$ as the coordinate transverse to the BPS curve. At fixed rotation it measures the displacement from the BPS curve,
\begin{equation}
\label{eq:epseta}
r_+^2-a(a+2)=(1+2a)\,\eta+\tfrac12\,\eta^2,\qquad \delta r_+\equiv r_+-\sqrt{a(a+2)}=\frac{(1+2a)\,\eta}{2\sqrt{a(a+2)}}+O(\eta^2).
\end{equation}
We express the analytic results in terms of $\eta$. The black-hole phase diagrams use $(r_+,a)$ coordinates, with $r_+$ horizontal; near the BPS curve, figures~\ref{fig:fixedn} and~\ref{fig:strip} resolve the displacement $\delta r_+$ or the indicated rescaling of it. In the $(r_+,a)$ plane, black holes with $\eta<0$ lie above and to the left of the BPS curve, in the range $r_+<\sqrt3$; those with $\eta>0$ lie below and to the right. The horizon angular velocity, measured in a frame that is nonrotating at infinity, and the electric potential are
\begin{equation}
\label{eq:OmPhi}
\begin{aligned}
&\Omega_H=\frac{a(1+s)}{\Theta_2},&\qquad &\Phi_H=\frac{s^2-s-2a^2}{\Theta_2},\\
&\Omega_H-1=-\frac{(1-a)\,\eta}{\Theta_2},&\qquad &\Phi_H-1=\frac{(s-1+2a)\,\eta}{\Theta_2},
\end{aligned}
\end{equation}
where $\Theta_2=s-1+2a^2>0$. Thus $\eta<0$, $\Omega_H>1$ and $\Phi_H<1$ are equivalent on $\Dom$; we call this region above the BPS curve and $\eta>0$ below it.

There are two anti-aligned families. The continuation of \eqref{eq:qm} beyond $q_C=0$, where $2r_+^4+r_+^2<a^2$, has $\Omega_H>1$ and $-1<\Phi_H<0$. The root $q_C=-a^2-sr_+^2$ has $\Omega_H<1$ and $\Phi_H<-1$, by \eqref{eq:OmPhi} with $s\to-s$. Neither is supersymmetric (appendix~\ref{app:antialigned}). Thus every non-BPS extremal equal-rotation black hole has $\Omega_H>1$ or $|\Phi_H|>1$. Section~\ref{sec:antialigned} proves BF violation throughout both families, so restricting the main analysis to $\Dom$ loses no BF-passing black hole.

\subsection{Spectrum and minimal coupling of the \texorpdfstring{$t^+$}{t+} Kaluza--Klein tower}\label{sec:tower}

The $t^+$ scalar spectrum follows from the Kaluza--Klein analysis~\mbox{\cite{Eloy:2026jzg}}. At level $n\in\mathbb Z_{\ge0}$ the scalars have charge and mass
\begin{equation}
\label{eq:tower}
Q=n+2,\qquad M^2=n^2-4=Q^2-4Q,
\end{equation}
and obey $(D^\mu D_\mu-M^2)\phi=0$ with $D=\nabla-iQ\AF$. At level $n$ they are the components of charge $\pm(n+2)$ in the $[0,n+2,0]$ of SU(4), which form the symmetric power $\operatorname{Sym}^{n+2}\mathbf3$ of SU(3) and its conjugate, twelve real fields at $n=0$. They are dual to half-BPS chiral primaries whose dimension equals the absolute value of their charge~\mbox{\cite{Witten:1998qj,Kim:1985ez,Gunaydin:1984fk}}. Charge conjugation maps $(Q,k)\to(-Q,-k)$ and leaves the bound of section~\ref{sec:bound} unchanged, so we may take $Q>0$ and all values of $k$.

The decoupling and vanishing Pauli coupling of this tower follow by specializing the general five-dimensional exceptional-field-theory equations to the identity scalar matrix and the minimal gauge embedding, as in the near-horizon application~\mbox{\cite{Eloy:2026jzg}}. Appendix~\ref{app:pauli} derives the level-zero cancellation from the $\mathrm{SL}(6)/\mathrm{SO}(6)$ action and extends it to every level: the charged scalar representation has no mixing partner at fixed internal harmonic degree, and the Pauli matrix acts only on its level-zero scalar factor. These covariant five-dimensional statements apply before separation into horizon harmonics. In the convention
\begin{equation}
\label{eq:pauliconvention}
\big(D^\mu D_\mu-M^2+\tfrac12\fp_n\mathcal F^2\big)\phi=0, \qquad \mathcal F=d\AF,
\end{equation}
the tower has $\fp_n=0$ \footnote{\label{fp}Ezroura and Larsen instead give $\fp_0=2$ for the level-zero $t^+$ fields~\mbox{\cite{Ezroura:2024xba}}, the difference is not removed by matching their effective-mass convention in appendix~\ref{app:pauli} .} for every $n\ge0$, with $\mathcal F^2=\mathcal F_{\mu\nu}\mathcal F^{\mu\nu}$.

\subsection{Horizon harmonics and the effective \texorpdfstring{AdS$_2$}{AdS2} scalar equation}\label{sec:nh}

In the extremal near-horizon limit, the CCLP geometry becomes a squashed three-sphere fibered over AdS$_2$~\mbox{\cite{Kunduri:2007vf,Eloy:2026jzg}}. For a black hole in $\Dom$, the near-horizon metric and gauge field take the form
\begin{equation}
\label{eq:nhmetric}
\begin{aligned}
ds^2&=-A^2x^2dt^2+B^2\frac{dx^2}{x^2}+\frac{C^2}{4}\big(\sigma_1^2+\sigma_2^2\big)+\frac{D^2}{4}\big(\sigma_3+E\,x\,dt\big)^2,\\
\AF&=-\big(F\,x\,dt+G\,\sigma_3\big).
\end{aligned}
\end{equation}
Here $x$ is the radial coordinate of AdS$_2$; $\sigma_i$ are the left-invariant one-forms on SU(2) in Euler angles $(\theta,\phi,\psi)$, with $\sigma_3=d\psi+\cos\theta\,d\phi$, and $A,\dots,G$ are functions of $(a,s)$ given in appendix~\ref{app:nhdata}. The Euler angles are related to the CCLP coordinates by
\begin{equation}
\label{eq:angles}
\theta=2\theta_C,\qquad \phi=\psi_C-\phi_C,\qquad \psi^0=\phi_C+\psi_C,\qquad \psi=\psi^0-2\Omega_Ht_C.
\end{equation}
Thus $\psi$ co-rotates with the horizon. We take $r-r_+=\epsilon x$ and $t_C=t/\epsilon$ as $\epsilon\to0$, after the gauge shift $\AF\to\AF-\Phi_Hdt_C$ removes the horizon potential. The horizon is a squashed three-sphere: a round two-sphere of radius $C/2$\footnote{\label{fn:C2}Eloy et al. give a near-horizon expression for $C^2$ with $1-a$ in its denominator~\mbox{\cite{Eloy:2026jzg}}. With $a=b$ the CCLP metric contains the base of the Hopf fibration as $\frac{r^2+a^2}{4(1-a^2)}(\sigma_1^2+\sigma_2^2)$, and this term does not rescale in the near-horizon limit. This gives \eqref{eq:C2} at $r=r_+$. This base-sphere radius agrees with the expression of Ezroura and Larsen~\mbox{\cite{Ezroura:2024xba}}. Since $C$ enters only through $j(j+1)-k^2$, this correction leaves the singlet results unchanged~\mbox{\cite{Eloy:2026jzg}}.}, with
\begin{equation}
\label{eq:C2}
C^2=\frac{r_+^2+a^2}{1-a^2}=\frac{s^2-1}{2(1-a^2)},
\end{equation}
and a Hopf fiber of radius $D$, twisted over AdS$_2$ by $E$ and threaded by the magnetic potential $-G\sigma_3$.

The gauge field is a globally well-defined one-form on the horizon. Charged fields are functions on $S^3$, and all its harmonics occur. We use the phase convention $D^j_{mk}(\theta,\phi,\psi)=e^{im\phi}d^j_{mk}(\theta)e^{ik\psi}$ for the Wigner functions, with $j\in\frac12\mathbb Z_{\ge0}$ and $m,k\in\{-j,\dots,j\}$. These functions are eigenfunctions of the SU(2) Casimir with eigenvalue $j(j+1)$ and of the fiber momentum operator $-i\partial_\psi$ with eigenvalue $k$; the label $m$ encodes a $(2j+1)$-fold degeneracy~\mbox{\cite{Murata:2008xr}}.\footnote{\label{fn:kphi}In terms of the CCLP azimuthal quantum numbers, $k=(m_\phi+m_\psi)/2$ and $m=(m_\psi-m_\phi)/2$. Singlets are the modes with $j=k=0$.} The singlet sector has already been reduced to a charged AdS$_2$ scalar~\mbox{\cite{Eloy:2026jzg}}. We extend this reduction to all horizon harmonics by inserting $\phi=e^{-i\omega t}f(x)D^j_{mk}$, obtaining a charged scalar on AdS$_2$ of radius $\ell_2=B$, with
\begin{equation}
\label{eq:m2q2}
m_2^2=M^2+\frac{4\,[\,j(j+1)-k^2\,]}{C^2}+\frac{4\,(k+QG)^2}{D^2},\qquad q_2=\frac{QF-E\,(k+QG)}{AB}.
\end{equation}
The angular terms are the energies on the base and fiber. The magnetic potential shifts $k$ to $k+QG$, while the fiber twist contributes to the AdS$_2$ charge. Neither effective parameter depends on $m$, so each mode $(n,j,k)$ retains the $2j+1$ horizon degeneracy and all internal components of charge $Q=n+2$.

\subsection{The near-horizon BF criterion and radial stability}\label{sec:bound}

At zero frequency in the throat, the AdS$_2$ equation has power-law solutions $x^{-1/2\pm\nu}$, with
\begin{equation}
\label{eq:beta}
4\nu^2=\beta\equiv1+4m_2^2\ell_2^2-4q_2^2\ell_2^4 .
\end{equation}
The exponents are real if and only if $\beta\ge0$, which is the charged BF bound in AdS$_2$~\mbox{\cite{PiolineTroost:2005AdS2}}. At $\beta=0$ they coincide, and the solutions are $x^{-1/2}$ and $x^{-1/2}\log x$. A mode satisfies the bound when $\beta\ge0$, saturates it when $\beta=0$ and violates it when $\beta<0$. A black hole is BF-passing if every $t^+$ mode satisfies the bound there, and BF-violating otherwise. We write
\begin{equation}
\label{eq:Sset}
S=\big\{(a,r_+)\in\Dom:\ \beta_{n,j,k}\ge0\ \text{for all }n,j,k\big\}
\end{equation}
for the BF-passing set. We label its points by either $(a,r_+)$ or $(a,\eta)$; in the latter coordinates, $(a,0)$ denotes a BPS black hole.

At nonzero throat frequency, the same exponents govern the AdS$_2$ boundary asymptotics rather than the Poincar\'e horizon~\mbox{\cite{GrallaZimmerman:2018Scaling}}. For a full mode with time dependence $e^{-i\omega t_C}$, the angular and gauge shifts above give throat frequency $\hat\omega=(\omega-\omega_s)/\epsilon$, where $\omega_s=2k\Omega_H-Q\Phi_H$. Thus the synchronized frequency $\omega=\omega_s$ gives zero throat frequency; a finite nonzero $\hat\omega$ requires $\omega-\omega_s=O(\epsilon)$.

Substituting the near-horizon data gives
\begin{equation}
\label{eq:betaclosed}
\begin{aligned}
\beta_{n,j,k}=1&+\frac{(s^2-1)M^2+8(1-a^2)j(j+1)}{\Theta_1}\\
&+\frac{2(1+s)(1+a^2-s+3a^2s)\,U^2Q^2-24a^2(1-a^2)(s^2-1)(1+s)^2\,k^2}{\Theta_1^2\Theta_2^2}\\
&-\frac{8a(1-a^2)(1+s)(1+4a^2+2s-3s^2)\,UQk}{\Theta_1^2\Theta_2^2},
\end{aligned}
\end{equation}
where
\begin{equation}
\Theta_1=3s^2-1-2a^2>0,\,\,\,U=s^2-s-2a^2=2q_C/(1+s)\ge0.
\end{equation}
For $j=k=0$, this expression reproduces the previously obtained singlet BF discriminant~\mbox{\cite{Eloy:2026jzg}}. The reduction is verified in appendix~\ref{app:closedform}.

At $k=0$ one has the identity
\begin{equation}
\label{eq:kzero}
\beta_{n,j,0}-\beta_{n,0,0}=\frac{8(1-a^2)j(j+1)}{\Theta_1}>0\qquad(j\ge1).
\end{equation}
Consequently, every $k=0$ violation already has a violating singlet at the same level.

On the Reissner--Nordstr\"om axis $a=0$, the fiber label drops out:
\begin{equation}
\label{eq:RNj}
\beta_{n,j,k}\big|_{a=0}=-3+\frac{\big[r_+^2(n-4)-2\big]^2}{(1+3r_+^2)^2}+\frac{4j(j+1)}{1+3r_+^2}.
\end{equation}
Each level is therefore minimized by its singlet.

This is a local criterion. For the charged, generally non-axisymmetric modes considered here, BF violation alone does not establish a growing radial mode~\mbox{\cite{Durkee:2010ea}}. For example, non-axisymmetric perturbations of the extremal Kerr throat can violate the effective BF bound, with their growth depending on the boundary condition at the edge of the throat~\mbox{\cite{Amsel:2009ev,Dias:2009ex}}. The Durkee--Reall vacuum conjecture assumes the relevant rotational symmetry, here $k=0$, and its scalar argument requires the absence of first time derivatives~\mbox{\cite{Durkee:2010ea}}. Neither applies to our general charged harmonics, and by \eqref{eq:kzero} every violating mode in a singlet-passing region has $k\ne0$. The vacuum gravitational and test-Maxwell results likewise do not supply a general implication for this charged scalar tower~\mbox{\cite{Hollands:2014lra,Green:2015kur}}. Conversely, passing the bound does not prove radial stability~\mbox{\cite{Dias:2010ma,Basu:2010uz}}.

At synchronization, $\omega=\omega_s$, the full radial equation has the same horizon exponents. In $X=r^2$ it is $(\Hor R')'+V_{\rm eff}R=0$, with the effective potential $V_{\rm eff}$ of \eqref{eq:Veff} and
\begin{equation}
\label{eq:Hor}
\Hor=(X-X_+)^2(X+1+2a^2+2X_+),\qquad X_+=r_+^2.
\end{equation}
The finite horizon limit $V_0$ obeys
\begin{equation}
\label{eq:horexp}
1-\frac{8V_0}{\Theta_1}=\beta,
\end{equation}
giving $R\sim(X-X_+)^{-1/2\pm\sqrt\beta/2}$ on both charge roots (appendix~\ref{app:horexp}).

\section{The BF parameter as a quadratic form in charge and spin}\label{sec:form}

On the domain $\Dom$, minimization over the fiber momentum reduces $\beta$ to a quadratic polynomial in charge and spin.

\subsection{Fiber momentum minimization and the signature of the quadratic form}\label{sec:kmin}

At fixed $(n,j)$, the closed form \eqref{eq:betaclosed} is a concave quadratic function of $k$ (appendix~\ref{app:kmin}). Its minimum over $|k|\le j$ is therefore attained at $k=\pm j$. The part of \eqref{eq:betaclosed} that is odd in $k$ is linear in $Qk$, and the two boundary values differ by
\begin{equation}
\label{eq:kdiff}
\begin{aligned}
\beta_{n,j,j}-\beta_{n,j,-j}&=16\,jQ\,B^2\Big[\frac{4G}{D^2}+\frac{E\,(F-EG)}{A^2}\Big]\\
&=\frac{16a(1-a^2)\,jQ\,(1+s)\,[(s^2-1)+2U]\,U}{\Theta_1^2\Theta_2^2}\ge0 ,
\end{aligned}
\end{equation}
hence, on all of $\Dom$,
\begin{equation}
\min_k\beta_{n,j,k}=\beta_{n,j,-j}.
\end{equation}
The minimizer is unique for $j>0$ and $U>0$; the endpoints tie at $U=0$. The harmonic $k=-j$ maximizes $|\omega_s|$ in the multiplet, with $\omega_s=-(2j\Omega_H+Q\Phi_H)\le0$. Its charge conjugate, obtained by $(\omega,Q,k)\to(-\omega,-Q,-k)$, has $k=j$ and synchronized frequency $2j\Omega_H+Q\Phi_H\ge0$; it co-rotates and is co-charged with the black hole, the orientation favored by superradiance. At the black holes with $q_C<0$ of section~\ref{sec:antialigned}, the difference \eqref{eq:kdiff} is negative instead, and the minimizing harmonic is $k=+j$.

\phantomsection\label{sec:det}
Setting $k=-j$ and using $M^2=Q^2-4Q$, we find that $\beta$ is an exact quadratic polynomial in $(Q,j)$,
\begin{equation}
\label{eq:form}
\beta_{n,j,-j}=\aQQ Q^2+2\aQj Qj+\ajj j^2+\aQ Q+\aj j+1,
\end{equation}
with coefficients
\begin{equation}
\label{eq:coeffs}
\begin{gathered}
\aQQ=\frac{(1+s)^2\NQQ}{\Theta_1^2\Theta_2^2},\qquad \aQj=-\frac{4a(1-a^2)(1+s)(3s^2-2s-1-4a^2)\,U}{\Theta_1^2\Theta_2^2},\\
\ajj=\frac{8(1-a^2)^2\Njj}{\Theta_1^2\Theta_2^2},\qquad \aQ=-\frac{4(s^2-1)}{\Theta_1},\qquad \aj=\frac{8(1-a^2)}{\Theta_1},
\end{gathered}
\end{equation}
where $\NQQ$ is defined in \eqref{eq:NQQ} and $\Njj$ in \eqref{eq:Njjeta}. Because
\begin{equation}
3s^2-2s-1-4a^2=(s^2-1)+2U>0,
\end{equation}
the mixed coefficient satisfies $\aQj\le0$, with equality only on the boundary $U=0$. The coefficient of $Q^2$ is positive on $\Dom$,
\begin{equation}
\label{eq:aQQpos}
\aQQ>0 ,
\end{equation}
as the parametrization in appendix~\ref{app:kmin} shows. In the variables $(a,\eta)$, the numerator $\Njj$ has the positive-coefficient expansion \eqref{eq:Njjeta}. Hence $\ajj>0$ on and below the BPS curve.

The sign structure of the form is controlled by the determinant of its quadratic part:
\begin{equation}
\label{eq:Delta}
\begin{aligned}
\Delta\equiv\aQQ\ajj-\aQj^2 &=\frac{8(1-a^2)^2(s^2-1)^2(s-1+2a)}{\Theta_1^3\Theta_2^2}\,\eta\\[2pt]
&=\frac{8(1-a^2)^2(s^2-1)^2}{\Theta_1^3\Theta_2}\,(\Phi_H-1)\\[2pt]
&=-\frac{8(1-a)(1+a)^2(s^2-1)^2(s-1+2a)}{\Theta_1^3\Theta_2}\,(\Omega_H-1).
\end{aligned}
\end{equation}
All factors other than $\eta$, $\Phi_H-1$ and $1-\Omega_H$ are positive. Therefore, below the BPS curve ($\eta>0$), the quadratic part is positive definite; above the BPS curve ($\eta<0$), it has signature $(1,1)$; on the BPS curve ($\eta=0$), it is a perfect square,
\begin{equation}
\label{eq:perfectsquare}
\aQQ Q^2+2\aQj Qj+\ajj j^2\Big|_{\eta=0} =\frac{4\big(2aQ-(1-a)j\big)^2}{(1+5a)^2},
\end{equation}
with null direction $j/Q=2a/(1-a)$.

\subsection{Large quantum numbers and spin minimization at fixed level}\label{sec:largelabels}

The signs in \eqref{eq:Delta} determine the behavior of $\beta$ at large quantum numbers. We call the scaling
\begin{equation}
(Q,j,k)=\Lambda(\bar Q,\bar\jmath,\bar k),\qquad \Lambda\to\infty,
\end{equation}
the large-quantum-number, or eikonal, limit. In this limit $M^2/Q^2\to1$, because $M^2=Q^2-4Q$, and $\Lambda^{-2}\beta$, minimized over $|\bar k|\le\bar\jmath$, tends to the quadratic part of \eqref{eq:form}.

Below the BPS curve, this quadratic part is positive definite. Large quantum numbers therefore cannot make $\beta$ arbitrarily negative. Above the BPS curve, where $\Delta<0$, the situation is different. When $\ajj>0$, the ratios $\bar\jmath/\bar Q$ for which the quadratic part is negative fill the open cone
\begin{equation}
\label{eq:cone}
t_-<\frac{\bar\jmath}{\bar Q}<t_+, \qquad t_\pm=\frac{-\aQj\pm\sqrt{-\Delta}}{\ajj}>0 .
\end{equation}
The roots are positive because $\aQQ>0$ and $\aQj<0$, so the cone lies inside the physical quadrant. If $\ajj<0$, there is a single positive root $t_-$, and the quadratic part is negative for $\bar\jmath/\bar Q>t_-$. If $\ajj=0$, it is negative for $\bar\jmath/\bar Q>\aQQ/(-2\aQj)$.

Since $\aQQ>0$, unbounded violation requires unbounded spin; minimizing sequences can be chosen at $k=-j$. As the BPS curve is approached, the cone closes onto $j/Q=2a/(1-a)$.

The sign of $\ajj$ also has a geometric interpretation. By \eqref{eq:ergo}, the horizon generator is spacelike in the throat precisely where $\ajj<0$; there, every fixed level has $\beta\to-\infty$ as $j\to\infty$ at $k=-j$. Where $\ajj>0$ above the BPS curve, unbounded violation requires both charge and spin to grow. On the boundary $\ajj=0$, the linear coefficient in spin determines the behavior at fixed level (section~\ref{sec:above}).

The limits of unbounded level or horizon momentum apply to the two-derivative supergravity operator. At finite string length, the momenta must remain below the string scale; that cutoff is not determined here.

\begin{figure}[tbp]
\centering
\includegraphics[width=\textwidth]{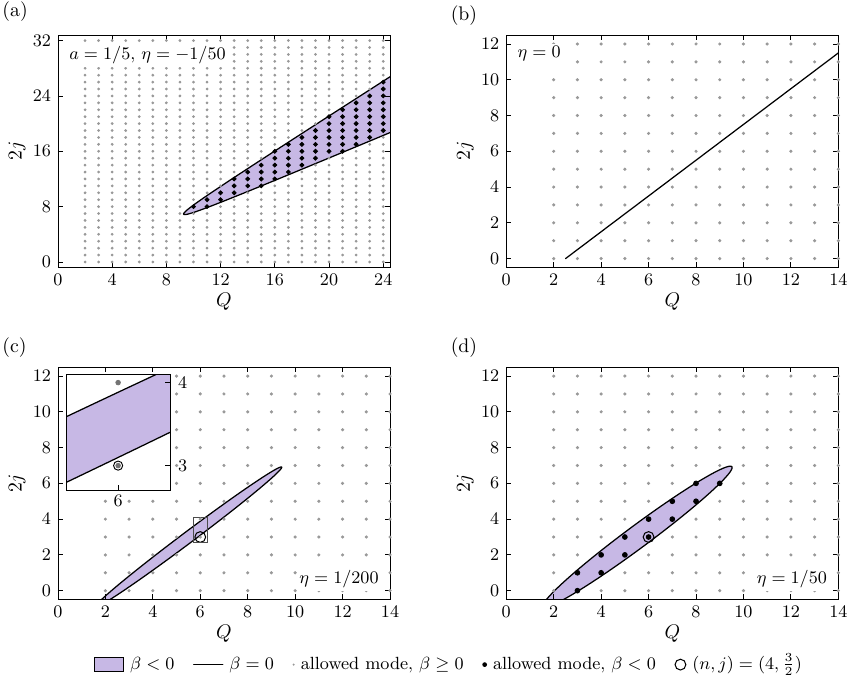}
\caption{The BF discriminant $\beta_{Q-2,j,-j}$ at $a=1/5$, plotted against charge $Q$ and twice the spin $2j$. Physical modes lie at $Q\in\mathbb Z_{\ge2}$ and $2j\in\Zp$: black dots violate the bound, while small gray markers satisfy it. Purple shading denotes the negative region of the same polynomial extended to real labels; black curves denote $\beta=0$. (a) Above the BPS curve, at $\eta=-1/50$, the negative region extends to arbitrarily large quantum numbers and contains infinitely many modes. (b) At $\eta=0$, $\beta$ is a perfect square. Its zero line misses every lattice point. (c) At $\eta=1/200$, the negative region is an ellipse containing no physical mode. The inset resolves its near miss of the circled mode $(n,j)=(4,\frac32)$ and obscures some gray markers in the main panel. (d) At $\eta=1/50$, this mode and several others lie inside the ellipse. The contrast between (c) and (d) illustrates why a negative continuous minimum need not imply BF violation by an allowed mode.}
\label{fig:plane}
\end{figure}

\phantomsection\label{sec:lowestmode}
The same coefficients determine which harmonic minimizes a given level at finite quantum numbers. At fixed $n$, the minimizing harmonic follows by rounding the vertex of
\begin{equation}
\label{eq:quadj}
\beta_{n,j,-j}=\ajj j^2+\lambda_n j+\beta_{n,0,0},\qquad \lambda_n=2\aQj Q+\aj.
\end{equation}
Where $\ajj>0$, define
\begin{equation}
\label{eq:spinsquare}
\beta_{n,j,-j}=\mu_n+\frac{\ajj}{4}(2j-y_n)^2,\qquad y_n=-\frac{\lambda_n}{\ajj},\quad \mu_n=\beta_{n,0,0}-\frac{\lambda_n^2}{4\ajj}.
\end{equation}
The level minimum is then
\begin{equation}
\label{eq:levelmin}
H_n\equiv\inf_{j,k}\beta_{n,j,k} =\mu_n+\frac{\ajj}{4}\dist(y_n,\Zp)^2,\qquad 2\hj_n=\max\{0,\lceil y_n-\tfrac12\rceil\},\quad k=-\hj_n.
\end{equation}
At a tie either adjacent spin minimizes. Since $\partial_Qy_n=-2\aQj/\ajj\ge0$, $\hj_n$ is non-decreasing with the level. For $q_C>0$, a spin-$\frac12$ mode lies below the singlet at sufficiently large $n$; where $\ajj>0$, $\hj_n/n\to-\aQj/\ajj>0$ (appendix~\ref{app:kmin}). Thus the singlet truncation misses the lowest harmonic of sufficiently high levels.

\section{BF bounds at fixed Kaluza--Klein level near the BPS curve}\label{sec:susy}

\subsection{The BF bound on the BPS curve and its saturation points}\label{sec:sos}

On the BPS curve, $\eta=0$, the factors entering \eqref{eq:betaclosed} reduce to
\begin{equation}
\label{eq:bpsD}
\Theta_1=2(1+a)(1+5a),\qquad \Theta_2=U=2a(1+a),\qquad s^2-1=4a(1+a),
\end{equation}
and $\beta$ can be written as a sum of squares. Collecting powers of $k$ gives, for every mode,
\begin{equation}
\label{eq:sos}
\beta^{\rm BPS}_{n,j,k}=\big(\Rch-1\big)^2 +\frac{4(1-a)}{1+5a}\,(j+k)(j-k+1), \qquad \Rch\equiv\frac{4aQ+2(1-a)k}{1+5a}.
\end{equation}
The second term is nonnegative and vanishes only at $k=-j$. Hence every BPS mode obeys $\beta\ge0$, with equality exactly when $k=-j$ and $\Rch=1$. This uses the zero Pauli coupling derived in appendix~\ref{app:pauli}; a positive Pauli term would lift every BPS mode above the bound and remove the saturation points, since $\ell_2^2\mathcal F^2=-2/(1+5a)<0$ on the BPS curve by \eqref{eq:Fsquared}.

\phantomsection\label{sec:tangency}
The condition $\Rch=1$ at $k=-j$ fixes the rotation in terms of $(n,j)$. The mode $(n,j,-j)$ saturates the bound at
\begin{equation}
\label{eq:ac}
a_c(n,j)=\frac{2j+1}{4n+2j+3}.
\end{equation}
For $j=0$ this gives the previously obtained singlet points $a_c=1/(4n+3)$~\mbox{\cite{Eloy:2026jzg}}. For $n=0$ it gives $a_c=(2j+1)/(2j+3)$.

Define
\begin{equation}
\label{eq:zeta}
c(a)=\frac{1-a}{1+5a}, \qquad \zeta_n(a)=\frac{(4n+3)a-1}{1-a}.
\end{equation}
For the minimizing endpoint harmonics $k=-j$, the BPS values of $\beta$ simplify to
\begin{equation}
\label{eq:alongcurve}
\beta^{\rm BPS}_{n,j,-j} =\frac{(4n+2j+3)^2(a-a_c)^2}{(1+5a)^2} =c(a)^2\big(\zeta_n-2j\big)^2.
\end{equation}
On the BPS curve, $y_n=\zeta_n$, so adjacent half-integer spins sample this square at spacing $c(a)$.

\subsection{Onset of BF violation above and below the BPS curve}\label{sec:leaving}

We now determine on which side of the BPS curve a saturating mode first develops $\beta<0$. Instead of completing the square in the spin, we complete it in the level:
\begin{equation}
\label{eq:levelsquare}
\beta_{n,j,-j} =\aQQ\,(n-v_j)^2+\eta\,\GG(j;a,\eta), \qquad v_j=-\frac{2\aQj j+\aQ}{2\aQQ}-2.
\end{equation}
Here $\GG$ is quadratic in $j$, independent of $n$, and regular at $\eta=0$; its leading coefficient is positive (appendix~\ref{app:sos}). On the BPS curve, the first term has a double zero at $v_j=n$, i.e.\ at $a=a_c(n,j)$. Writing $u=a-a_c$, we therefore expand near the saturation point as
\begin{equation}
\label{eq:tangexp}
\begin{gathered}
\beta_{n,j,-j} =K_{n,j}\,u^2+\Sigma_{n,j}\,\eta +O\big(|u|^3+|u|\,|\eta|+\eta^2\big),\\[2pt]
K_{n,j}=\frac{(4n+2j+3)^4}{16\,(n+3j+2)^2}, \qquad u=a-a_c.
\end{gathered}
\end{equation}
The curvature $K_{n,j}$ follows from \eqref{eq:alongcurve}. The linear coefficient is
\begin{equation}
\label{eq:P}
\begin{gathered}
\Sigma_{n,j}=\GG(j;a_c,0) =\frac{(4n+2j+3)\,P(n,j)}{4\,(n+j+1)(n+3j+2)},\\[2pt]
P(n,j)=2j\,(n^2-3n-1)-(3n^2+7n+2).
\end{gathered}
\end{equation}
When $P(n,j)<0$, the mode has $\beta<0$ just below the BPS curve, where $\Phi_H>1$. When $P(n,j)>0$, it has $\beta<0$ just above the BPS curve, where $\Omega_H>1$. In both cases $\Sigma_{n,j}\neq0$, so near the saturation point the zero set is the parabola
\begin{equation}
\eta=-\frac{K_{n,j}}{\Sigma_{n,j}}\,u^2+O(u^3),
\end{equation}
which is tangent to the BPS curve; the region with $\beta<0$ lies on the side of this parabola away from the BPS curve. In this nondegenerate case we call $a_c(n,j)$ a tangency point and say that the mode violates the bound first below or first above the BPS curve.

To compare levels at fixed rotation, write
\begin{equation}
\label{eq:Pa}
\mathcal P(a,Q)=a(2Q^2-12Q+15)-(2Q-3),\qquad P(n,j)=\tfrac12(4n+2j+3)\mathcal P(a_c,n+2).
\end{equation}
The positive quadratic term dominates at high charge and gives first violation above the curve; the lower-order terms reverse the sign at low charge. By \eqref{eq:Pderiv}, $2\mathcal P(a,Q)/[(1+a)(1+5a)]$ is the slope of the level minimum over real spin at $\eta=0$. Its coefficient of $Q^2$ is
\begin{equation}
\label{eq:tauhat}
\tauh(a)=\left.\frac{\Delta}{\eta\ajj}\right|_{\eta=0} =\frac{4a}{(1+a)(1+5a)}.
\end{equation}

For $n\ge3$, the function $\mathcal P(a,n+2)=a(2n^2-4n-1)-(2n+1)$ changes sign at
\begin{equation}
\label{eq:astar}
a_*(n)=\frac{2n+1}{2n^2-4n-1},\qquad n\ge3,
\end{equation}
and $a_*(n)<1/(n-\tfrac52)$. A saturating mode of level $n\ge3$ therefore violates the bound first below the BPS curve when $a_c<a_*(n)$, and first above it when $a_c>a_*(n)$.

The classification is now straightforward. For $n\le2$, $\mathcal P$ is negative. Since $a_*(3)=\frac75>1$, the saturating modes of levels $0$--$3$ all violate the bound first below the BPS curve. For larger $n$, $a_*(n)$ decreases, beginning with $a_*(4)=\frac35$. At a BPS black hole of rotation $a$, the saturating modes of levels satisfying $a_*(n)>a$---and hence satisfying the necessary bound $n<1/a+\frac52$---violate the bound first below the curve; those with $a_*(n)<a$ do so first above it. In terms of the spin labels,
\begin{equation}
\label{eq:Tn}
P(n,j)=(n^2-3n-1)\,(2j-T_n), \qquad T_n=3+\frac{16n+5}{n^2-3n-1}.
\end{equation}
For each level $n\ge4$, only finitely many spins violate the bound first below the BPS curve; they satisfy $2j<T_n$, with $T_n\le26$. From $n=20$ onward, only the four sequences $j\le\frac32$ remain. These sequences violate the bound first below the BPS curve at every level, and their tangency points accumulate at $a=0$. Thus at a fixed rotation $0<a<1$, every mode of sufficiently high level that saturates there first violates the bound above the curve, but no level threshold holds uniformly in $a$. In particular, each singlet has $P(n,0)=-(3n^2+7n+2)<0$, so its BF-violating region lies below the BPS curve near the tangency point $1/(4n+3)$.

The only exceptional mode is $(n,j)=(4,13)$, which saturates at $a=\frac35$ with $P(4,13)=0$. Its saturation point is a nondegenerate saddle: the nearby zero set consists of two crossing curves, and at fixed $a=\frac35$ the mode violates the bound on both sides of the BPS curve. The mode $(0,1,-1)$ saturates at the same rotation and has $P(0,1)=-4$.

\subsection{BF boundaries at fixed level and isolated violating regions}\label{sec:levels}\label{sec:neartangency}

\begin{figure}[tbp]
\centering
\includegraphics[width=\textwidth]{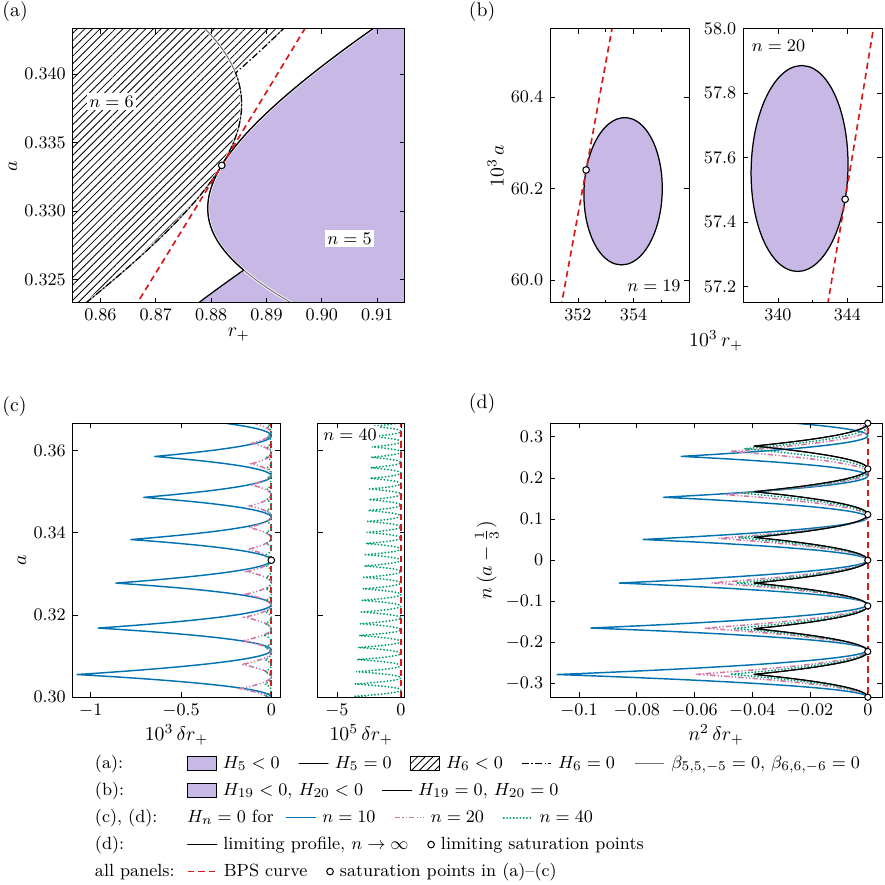}
\caption{Fixed-level minima $H_n$ near the BPS curve (red dashed). (a) In the $(r_+,a)$ plane, purple shading gives $H_5<0$ and hatching gives $H_6<0$, with solid and dash-dotted black boundaries respectively. Thin gray curves show the zero sets of the individual saturating modes $(5,5,-5)$ and $(6,6,-6)$. (b) The isolated violating regions at levels 19 and 20 lie on opposite sides of the BPS curve; both axes are multiplied by $10^3$, and $(j,k)=(2,-2)$ minimizes each level within its island. (c) The boundaries $H_n=0$ are blue solid for $n=10$, pink dash-dotted for $n=20$ and green dotted for $n=40$. For each level the bound is satisfied to the right of its curve. The main horizontal axis is $10^3\delta r_+$; the side panel resolves level 40 using $10^5\delta r_+$ with the same vertical scale. (d) Rescaling to horizontal coordinate $n^2\delta r_+$ and vertical coordinate $n(a-\frac13)$ displays convergence to the black limiting profile $n^2\delta r_+=-\frac{5\sqrt7}{84}\dist\big(9n(a-\frac13),\mathbb Z\big)^2$. Open circles mark saturation points in (a)--(c) and the limiting points $n(a-\frac13)\in\mathbb Z/9$ in (d).}

\label{fig:fixedn}
\end{figure}

Minimizing over all harmonics changes the single-level regions previously found from the singlets~\mbox{\cite{Eloy:2026jzg}}. Figure~\ref{fig:fixedn} shows the resulting saturation points, kinks and high-level limit; appendix~\ref{app:neartangency} supplies the proofs.

Near a saturation point, the pair $(j,k)=(j,-j)$ uniquely minimizes its level, retaining the $2j+1$ degeneracy in $m$, and $H_n=\beta_{n,j,-j}$ locally. For $P(n,j)\ne0$, converting \eqref{eq:tangexp} to the radial displacement gives
\begin{equation}
\label{eq:tangphys}
\begin{gathered}
H_n=K_{n,j}\,u^2+\frac{2r_c\,\Sigma_{n,j}}{1+2a_c}\,\delta r_++O\big(|u|^3+|u|\,|\delta r_+|+(\delta r_+)^2\big),\\[2pt]
r_c=\sqrt{a_c(a_c+2)},\qquad u=a-a_c.
\end{gathered}
\end{equation}
The zero set is therefore
\begin{equation}
\delta r_+\big|_{H_n=0} =-\frac{(1+2a_c)\,K_{n,j}}{2r_c\,\Sigma_{n,j}}\,u^2+O(u^3).
\end{equation}
The sign of $P$ fixes the violating side as in section~\ref{sec:leaving}; the exceptional mode $(4,13)$ instead has the crossing zero set described there. Away from saturation, adjacent spins tie at $y_n=2j+\frac12$, producing the kinks visible in figure~\ref{fig:fixedn}.

For example, $a_c(n,n)=\frac13$ and $P(n,n)=(2n+1)(n^2-5n-2)$. At this rotation, the local fixed-level expansions $H_5\simeq-3\sqrt7\,\delta r_+/10$ and $H_6\simeq3\sqrt7\,\delta r_+/5$ show that the first violating side changes between levels 5 and 6.

Figure~\ref{fig:fixedn}(b) illustrates the change of violating side between levels 19 and 20. For $j=2$, $a_c(n,2)=5/(4n+7)$ and $P(n,2)=n^2-19n-6$, so level 19 first violates below the BPS curve and level 20 above it. Each has a bounded BF-violating component whose closure meets the BPS curve only at its saturation point; we call such a component an island. In both islands $(j,k)=(2,-2)$ minimizes the level, and each island is convex in $(a,\eta)$.

\subsection{Scaling of the BF boundaries at large level}\label{sec:highdents}

Near a fixed interior rotation $a_0$, the tangency points become dense as the level increases. Successive tangency points are separated by approximately $(1-a_0)^2/(4n)$ in $a$, and all of them violate the bound first above the BPS curve, because
\begin{equation}
a_c=a_0+O(1/n), \qquad a_*(n)<\frac{1}{n-\frac52}
\end{equation}
by \eqref{eq:astar}. To resolve the structure between them, let $2j_n$ be an integer nearest to $\zeta_n(a_0)$, set
\begin{equation}
a_n=a_c(n,j_n), \qquad a=a_n+\frac{\hat u}{n}, \qquad \eta=\frac{\hat v}{n^2}.
\end{equation}
With $c_0=c(a_0)$ and $\tauh_0=\tauh(a_0)$, we find
\begin{equation}
\label{eq:profile}
H_n =c_0^2\,\dist\left(\frac{4\hat u}{(1-a_0)^2},\mathbb Z\right)^2 +\tauh_0\,\hat v+O(1/n),
\end{equation}
for fixed $a_0\in(0,1)$ and bounded scaled variables; appendix~\ref{app:profile} gives the precise uniformity. The first term comes from $H_n^{\rm BPS}=c(a)^2\dist(\zeta_n,\Zp)^2$; the second has the sign of $1-\Omega_H$. The two terms balance on the profile
\begin{equation}
\label{eq:vinf}
\hat v_\infty(\hat u) =-\frac{c_0^2}{\tauh_0}\, \dist\left(\frac{4\hat u}{(1-a_0)^2},\mathbb Z\right)^2\le0.
\end{equation}
The limiting region in which level $n$ satisfies the bound is $\hat v\ge\hat v_\infty(\hat u)$, to the right of the profile in figure~\ref{fig:fixedn}(d). For large $n$, the finite-level boundary is also a graph, which converges to this one with error $O(1/n)$ on compact intervals of $\hat u$.

Between consecutive tangency points, the zero set of $H_n$ and the BPS curve enclose an above-BPS region in which level $n$ satisfies the bound. We call this region a dent of level $n$. For large $n$, the deepest point of a dent is the kink at which two adjacent spins tie. At that point both spins saturate the bound, and
\begin{equation}
\eta_{\rm dent} =-\frac{c_0^2}{4\tauh_0n^2}+O(n^{-3}) =-\frac{(1-a_0)^2(1+a_0)}{16a_0(1+5a_0)n^2} +O(1/n^3).
\end{equation}
Thus the dents of a high level have width $O(1/n)$ in $a$ and depth $O(1/n^2)$. At $a_0=\frac13$,
\begin{equation}
\eta_{\rm dent}=-\frac{1}{24n^2}+O(n^{-3}), \qquad \delta r_{+,\rm dent}=-\frac{5\sqrt7}{336n^2}+O(n^{-3}).
\end{equation}

Finite-level corrections are visible in figure~\ref{fig:fixedn}(c,d): by \eqref{eq:Pderiv}, the slope in $\eta$ at $a=\frac13$ is proportional to $\mathcal P(\frac13,n+2)=\frac23(n^2-5n-2)$ rather than to its leading term $\frac23n^2$. The profile and the exhaustion by dents hold uniformly on compact rotation intervals; they do not describe the fixed-spin sequences $j\le\frac32$ accumulating at $a=0$.

\section{The BF bound for the full \texorpdfstring{$t^+$}{t+} Kaluza--Klein tower}\label{sec:wholetower}

We now minimize over the whole tower. By \eqref{eq:Delta}, the quadratic part of \eqref{eq:form} is indefinite above the BPS curve and positive definite below it, so the two sides behave differently: above the curve the violation is unbounded, while below it the negative region is an ellipse and the bound is decided by whether it contains a point of the charge--spin lattice.

\subsection{Unbounded BF violation above the BPS curve}\label{sec:above}

For $\eta<0$, where $\Omega_H>1$, the singlet analysis leaves a region in which every singlet satisfies the bound~\mbox{\cite{Eloy:2026jzg}}. Once all harmonics are included, every black hole above the BPS curve violates the bound and $\beta$ is unbounded below over the tower:
\begin{equation}
\label{eq:aboveempty}
S\cap\{\Omega_H>1\}=\emptyset, \qquad \inf_{n,j,k}\beta_{n,j,k}=-\infty .
\end{equation}

This follows from $\Delta<0$. If $\ajj<0$, as on the edge $U=0$ where $\Njj=-(s-1)^2(s+1)^2$, the spin quadratic \eqref{eq:quadj} is unbounded below at every level; this case occurs exactly where the horizon generator is spacelike in the throat (appendix~\ref{app:ergo}). Otherwise $U>0$ and $\aQj<0$, by \eqref{eq:coeffs}. If $\ajj>0$, minimizing over real spin gives
\begin{equation}
\label{eq:muQ}
\mu_n=\frac{\Delta}{\ajj}Q^2+\left(\aQ-\frac{\aQj\aj}{\ajj}\right)Q+1-\frac{\aj^2}{4\ajj}.
\end{equation}
For $Q\ge\aj/(-2\aQj)$ the vertex $y_n$ is nonnegative, and rounding yields
\begin{equation}
\label{eq:roundbound}
H_n\le\bar\mu_n\equiv\mu_n+\frac{\ajj}{16}\longrightarrow-\infty.
\end{equation}
If $\ajj=0$, the spin quadratic is linear with negative slope once $Q>\aj/(-2\aQj)$. In all three cases, minimizing harmonics of large spin make $\beta$ arbitrarily negative. This behavior was found for uncharged scalars on the edge $q_C=0$ of the family~\mbox{\cite{Durkee:2010ea}}. It is also consistent with the rotational superradiant threshold $\Omega_H=1$~\mbox{\cite{Green:2015kur}}.

The violating modes supplied by \eqref{eq:roundbound} may have large quantum numbers, and near some rotations they must. Let $a_0$ be a rotation at which no mode saturates the bound on the BPS curve; an irrational rotation provides an example. Given any $N,J\ge0$, there exists $\delta>0$ such that, for every black hole $(a_0,\eta)$ with $-\delta<\eta<0$,
\begin{equation}
\beta_{n,j,k}(a_0,\eta)\ge0
\end{equation}
for all modes with $n\le N$ and all modes with $j\le J$ (appendix~\ref{app:above}). Consequently, no finite truncation of the tower in level or spin detects the violation at every black hole above the BPS curve. As the required quantum numbers grow, the string-scale qualification of section~\ref{sec:largelabels} becomes essential.

\begin{figure}[tbp]
\centering
\includegraphics[width=\textwidth]{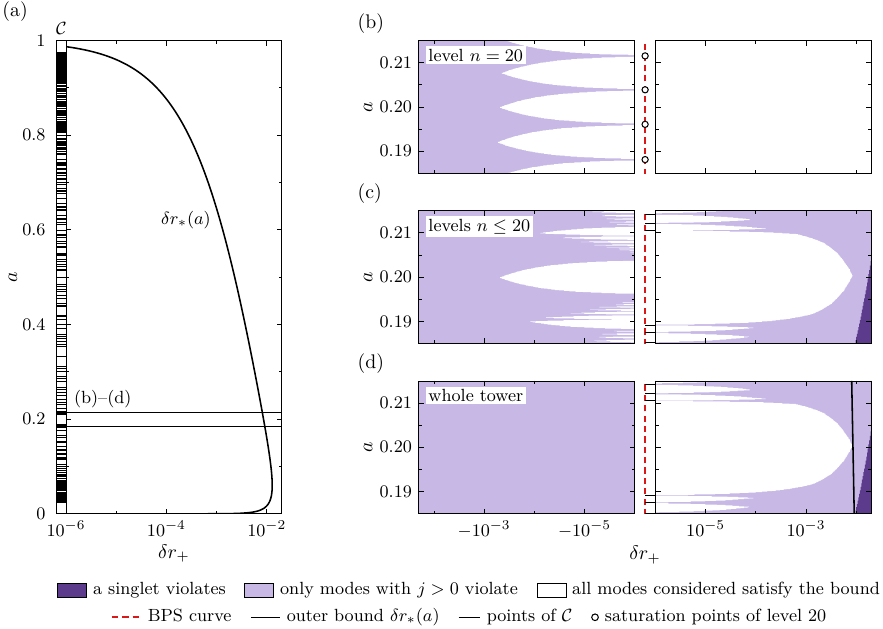}
\caption{Finite-level and whole-tower BF bounds near the BPS curve, with rotation $a$ vertical. (a) The black curve $\delta r_*(a)$ is the radial image of $\eta_*(a)$ under \eqref{eq:epseta}: every point to its right violates the bound, while points to its left require the lattice test. The separate bar represents the BPS curve $\delta r_+=0$; its ticks mark points of $\Cset$ with denominator at most 40. The rectangle selects the band $0.185\le a\le0.215$ enlarged in (b)--(d), comparing level 20, levels $0\le n\le20$, and the whole tower, each minimized over all harmonics. Their horizontal axes show signed $\delta r_+$, logarithmically in its magnitude, for $10^{-6}\le|\delta r_+|\le2\times10^{-2}$: negative values (left) are above the BPS curve and positive values (right) below it. The central break omits $|\delta r_+|<10^{-6}$; the red dashed line within it marks $\delta r_+=0$. White denotes satisfaction of the bound by every included mode, light purple violation only at nonzero spin, and dark purple singlet violation. Circles in (b) mark level-20 saturation points; ticks in (c) and (d) mark points of $\Cset$. The black outer bound is repeated in (d). The comparison shows how finite truncations retain BF-passing regions above the BPS curve that disappear when the complete tower is included.}

\label{fig:strip}
\end{figure}

At $a=\frac15$, no mode saturates on the BPS curve: $\zeta_n=n-\frac12$ and $H_n=1/25$ at every level. As $\eta\to0^-$, the first violating level and its minimizing harmonic satisfy
\begin{equation}
\label{eq:firstlevelmainasymptotic}
n_{\rm v}(\eta)=\frac{\sqrt3}{5\sqrt{|\eta|}}+O(1), \qquad j_{\rm v}=\frac{n_{\rm v}}2,\qquad k_{\rm v}=-j_{\rm v}.
\end{equation}
The first violating levels at $\eta=-10^{-2},-10^{-3},-10^{-4},-10^{-5}$ are respectively 9, 15, 38 and 113; they follow from an exact integer rule valid for $-1/100\le\eta<0$ (appendix~\ref{app:above}). Figure~\ref{fig:strip}(b)--(d) compares level 20, the levels $n\le20$, and the whole tower near this rotation.

Consequently, at $a=\frac15$ a truncation $n\le N$ leaves an above-BPS layer adjoining $\eta=0$ in which every retained mode satisfies the bound, although the complete tower contains a violating mode. Its width in $\eta$ is asymptotic to $3/(25N^2)$ as $N\to\infty$; for finite $N\ge9$, the boundary follows from the same integer rule. At any rotation without a saturating mode, both the internal level and the spin of the first violating mode diverge as the BPS curve is approached at finite horizon radius. At the saturation point of a mode $(n,j,-j)$ with $P(n,j)>0$, such as $(6,6,-6)$ at $a=\frac13$, that fixed mode instead violates the bound for all sufficiently small $\eta<0$ (section~\ref{sec:leaving}).

The large-radius limit of section~\ref{sec:large} involves a different scaling: the minimizing level remains $n=4$, while the spin grows with the horizon radius and the spin per unit horizon length remains finite at fixed rotation.

\subsection{The charge--spin lattice criterion below the BPS curve}\label{sec:below}\label{sec:ellipse}

Below the BPS curve, $\eta>0$ and $\Phi_H>1$, and the quadratic part of \eqref{eq:form} is positive definite. Let $(\Qmin,\jmin)$ denote the unique minimizer of \eqref{eq:form} over real $(Q,j)$, and let $\bmin$ be the corresponding minimum. Completing the square first in $j$ and then in $Q$, with
\begin{equation}
\rho=\frac{\aQj}{\ajj}\le0,
\end{equation}
we obtain
\begin{equation}
\label{eq:ellipse}
\beta_{n,j,-j} =\bmin+\frac{\Delta}{\ajj}\,(Q-\Qmin)^2 +\ajj\,\big(j-\jst(Q)\big)^2, \qquad \jst(Q)=\jmin-\rho\,(Q-\Qmin).
\end{equation}
Both curvatures are positive. As $Q+j\to\infty$ in the physical quadrant, $\beta\to+\infty$. Therefore only finitely many modes can violate the bound at a given black hole, and the minimum of $\beta$ over the tower is attained. In the variables $(a,\eta)$, the minimum has the form
\begin{equation}
\label{eq:bminform}
\bmin=-\frac{\eta\,N_\beta(a,\eta)}{(2a+\eta)^2(4a+\eta)\,\Theta_1},
\end{equation}
where $N_\beta$ is a polynomial with positive coefficients, given in \eqref{eq:bmin}, so $\bmin<0$ for $\eta>0$. Consequently, in the continuous relaxation every black hole below the BPS curve has some real $(Q,j)$ with $\beta<0$. Satisfaction of the BF bound is therefore a lattice effect: a black hole is BF-passing only if the open ellipse contains no point of $\mathbb Z_{\ge2}\times\frac12\mathbb Z_{\ge0}$.

The location of the center depends on $a$. If $\jmin<0$, which requires $a<(\sqrt{103}-7)/54$, then the true minimum over $j\ge0$ lies on the singlet line $j=0$, at
\begin{equation}
Q_0^*=-\frac{\aQ}{2\aQQ}>\Qmin>2,
\end{equation}
and this constrained minimum is also negative (appendix~\ref{app:below}). The lattice structure yields a finite decision procedure. Define
\begin{equation}
\label{eq:finitecharges}
\mathcal Q_-= \left\{ Q\in\mathbb Z_{\ge2}: |Q-\Qmin|<\sqrt{-\bmin\,\ajj/\Delta} \right\}.
\end{equation}
Then
\begin{equation}
\label{eq:finitecriterion}
(a,\eta)\in S \iff \min_{Q\in\mathcal Q_-}\mathcal B(Q)\ge0,
\end{equation}
where
\begin{equation}
\mathcal B(Q) =\bmin+\frac{\Delta}{\ajj}\,(Q-\Qmin)^2 +\ajj\,\dist\Big(\jst(Q),\tfrac12\mathbb Z_{\ge0}\Big)^2,
\end{equation}
and the minimum over an empty set is interpreted as $+\infty$. The ellipse itself determines the charges requiring inspection, so this criterion needs no universal cutoff in charge or spin. Figure~\ref{fig:plane}(c,d) shows the ellipse and the lattice at two black holes below the BPS curve: one BF-passing and one BF-violating.

The ellipse construction also computes a positive full minimum, using the closed charge interval determined by a trial mode in \eqref{eq:fullminimuminterval}. Section~\ref{sec:highlevels} gives a uniform twenty-four-minimum criterion for membership in $S$, which need not equal a positive full minimum.

The violation nevertheless remains bounded:
\begin{equation}
\beta_{n,j,k}>-3
\end{equation}
for every mode at every black hole in $\Dom$ on or below the BPS curve. This lower bound is sharp as an infimum; the large-black-hole analysis of section~\ref{sec:large} shows that the infimum equals $-3$. Together with sections~\ref{sec:sos} and~\ref{sec:above}, it follows that, for black holes in $\Dom$,
\begin{equation}
\inf_{n,j,k}\beta_{n,j,k}=-\infty
\end{equation}
if and only if $\Omega_H>1$.

The continuous minimum also bounds the strength of violation near the BPS curve. For every mode at every black hole in $\Dom$ with $\eta>0$,
\begin{equation}
\label{eq:linearstrength}
\beta_{n,j,k}\ge\bmin>-g(a)\eta, \qquad g(a)=\frac{1+6a+6a^2}{a(1+a)(1+5a)}.
\end{equation}
The coefficient is sharp for the continuous minimum: $-\bmin/\eta\to g(a)$ as $\eta\to0^+$, uniformly on compact subintervals of $0<a<1$. Attainment by allowed modes is a separate question, and the estimate is not uniform as $a\to0$.

\subsection{Satisfaction of the BF bound just below the BPS curve}\label{sec:towerdents}

To determine which black holes immediately below the BPS curve are BF-passing, we collect the rotations at which a saturating mode has $P\le0$:
\begin{equation}
\label{eq:Cset}
\Cset=\big\{a_c(n,j):\ P(n,j)\le0\big\}.
\end{equation}
In appendix~\ref{app:reach} we show that at every point of $\Cset$, some mode violates the bound for all sufficiently small $\eta>0$. The set $\Cset$ consists entirely of rational rotations. It has no accumulation points in $(0,1)$, accumulating only at $a=0$ through the sequences $j\le\frac32$ and at $a=1$ through the levels $n\le3$. It contains the singlet tangency points $1/(4n+3)$ and the level-zero tangency points $(1+2j)/(3+2j)$. These are the only obstructions immediately adjacent to the BPS curve. The BF-passing set contains a one-sided neighborhood
\begin{equation}
\{|a-a_0|<\delta,\ 0<\eta<\delta\}
\end{equation}
of a BPS rotation $a_0$ if and only if $a_0\notin\Cset$. The proof uses the completed-square form \eqref{eq:levelsquare}: every mode with sufficiently large spin has $\beta\ge\eta$, while for the remaining finitely many spins either $\GG(j;a,0)>0$ or the lattice point misses the ellipse, except when $a$ is a saturation point with $P\le0$. As a consequence, the measure of the set of rotations $a$ for which $(a,\eta)$ is BF-passing tends to one as $\eta\to0^+$.

Since $\Cset$ is locally finite, every interior BPS point $(a_0,0)$, including those with $a_0\in\Cset$, is a limit of points of $S$ with $\eta>0$. At $a_0\in\Cset$, such an approach must vary the rotation; no segment $\{a_0\}\times(0,\delta)$ lies in $S$. Each interval between consecutive singlet saturation points contains further points of $\Cset$, so the singlet dents are not contained in $S$.

The approach to full measure is nonuniform at the rotation endpoints. Writing $V_\eta=\{a\in(0,1):(a,\eta)\notin S\}$, on each compact interval $I\Subset(0,1)$ with endpoints outside $\Cset$ the excluded measure is $O_I(\sqrt\eta)$, with explicit weights given in \eqref{eq:measurelocal}. Near $a=0$, however, singlets alone exclude $0<a\le(\eta/8)^{1/3}$, yielding the global lower bound $|V_\eta|\ge\frac12\eta^{1/3}$ for $0<\eta\le1/125$ by \eqref{eq:measurelower}. Thus the interior formula does not determine the global leading asymptotic behavior.

\subsection{A strip enclosing the region satisfying the BF bound}\label{sec:width}

An outer bound on the BF-passing set follows by rounding the center of the ellipse. Let $Q_\circ$ be an integer nearest to $\Qmin$ and $j_\circ$ a half-integer nearest to $\jst(Q_\circ)$, initially without restricting its sign, so that
\begin{equation}
|Q_\circ-\Qmin|\le\tfrac12, \qquad \big|j_\circ-\jst(Q_\circ)\big|\le\tfrac14.
\end{equation}
Whenever $j_\circ\ge0$, this pair labels a physical mode and
\begin{equation}
\label{eq:Psi}
\beta_{Q_\circ-2,j_\circ,-j_\circ} \le \Psi(a,\eta) \equiv \bmin+\frac{\Delta}{4\ajj}+\frac{\ajj}{16}.
\end{equation}
The charge is always admissible since $\Qmin>2$. Appendix~\ref{app:width} proves spin admissibility outside the corner $a<\frac1{29}$, $\eta\ge12a$; in that corner a singlet supplies the violating mode instead. The numerator of $\Psi$ can be written as
\begin{equation}
\label{eq:NPsi}
N_\Psi =2048\,a^7(1-a)^2(1+a)^6 -\sum_{i=1}^{13}c_i(a)\,\eta^i,
\end{equation}
where each $c_i(a)$ is a nonzero polynomial with nonnegative coefficients. The denominator is $4(2a+\eta)^2(4a+\eta)\Theta_2^2\Theta_1^2\Njj>0$. Hence $N_\Psi$ decreases strictly with $\eta$ and has a unique positive zero $\eta_*(a)$. The inequality $\Psi<0$ is therefore equivalent to $\eta>\eta_*(a)$. With the singlet replacement in the exceptional corner, every black hole with $\eta>\eta_*(a)$ is BF-violating. Thus
\begin{equation}
S\cap\{\Omega_H<1\} \subset \{0<\eta\le\eta_*(a)\}.
\end{equation}
Exact evaluation and a Sturm count give the uniform outer bound
\begin{equation}
\label{eq:stripmaximum}
\eta_*(a)<0.0094\qquad(0<a<1).
\end{equation}
The same methods give a tighter bracket for its supremum, and many points on $\eta=\eta_*(a)$ already violate the bound. This curve therefore encloses $S$ rather than identifying its boundary; detached components inside the strip are not excluded.

The same outer strip bounds the depth of any violations that remain inside it:
\begin{equation}
\label{eq:stripfloor}
\beta_{n,j,k}>-\frac14 \qquad\text{for }0<a<1,\quad 0<\eta\le\eta_*(a).
\end{equation}
This holds for every mode, including at BF-violating black holes within the strip. The constant $\frac14$ is sharp for the continuous minimum, approached along $\eta=\eta_*(a)$ as $a\to0$; whether allowed modes approach $\beta=-\frac14$ is not determined.

\subsection{Uniform positivity and a BF criterion using twenty-four minima}\label{sec:highlevels}

For every $0<a<1$ and $\eta>0$,
\begin{equation}
\label{eq:highlevelpositive}
\beta_{n,j,k}>0\qquad(n\ge20,\ j\ge2,\ |k|\le j).
\end{equation}
Appendix~\ref{app:highlevels} proves this uniformly by exact polynomial inequalities. Both thresholds are sharp: $P(19,2)=-6$ and $P(20,\frac32)=-325$ give violation just below their respective saturation points. It remains to check levels $n\le19$ and the four fixed-spin sequences $j\le\frac32$.

Along each of the twenty levels $0\le n\le19$ and the four spins $j\in\{0,\frac12,1,\frac32\}$ at $n\ge20$, $\beta_{n,j,-j}$ is strictly convex in the remaining quantum number. The fixed-level minima are $H_n$ in \eqref{eq:levelmin}; for the four spins define
\begin{equation}
\label{eq:spinminima}
\hat n_j=\max\big\{20,\lceil v_j-\tfrac12\rceil\big\},\qquad \widetilde H_j=\beta_{\hat n_j,j,-j}=\aQQ(\hat n_j-v_j)^2+\eta\,\GG(j;a,\eta).
\end{equation}
Then
\begin{equation}
\label{eq:F24}
\begin{gathered}
F_{24}(a,\eta)=\min\{H_0,\dots,H_{19},\widetilde H_0,\widetilde H_{1/2},\widetilde H_1,\widetilde H_{3/2}\},\\
(a,\eta)\in S\quad\Longleftrightarrow\quad F_{24}(a,\eta)\ge0, \qquad 0<a<1,\quad\eta>0.
\end{gathered}
\end{equation}
Indeed, after minimizing over $k$, the omitted modes have $n\ge20$ and $j\ge2$, hence strictly positive $\beta$. Their minimum is attained because the quadratic form is positive definite. The full minimum has the sign of $F_{24}$ and equals it whenever $F_{24}\le0$. When $F_{24}>0$, an omitted mode may lie below it; thus this criterion does not in general compute a positive full minimum.\footnote{At $a=5/87$ on the BPS curve, $\beta_{n,j,-j}=(10n-36-82j)^2/56^2$ and the continuous extension of $F_{24}$ is $1/3136$, whereas $(20,2,-2)$ saturates. For sufficiently small $\eta>0$, this mode is strictly positive and smaller than $F_{24}$.} The twenty-four selected modes vary with the black hole and need not have bounded quantum numbers near $a=0,1$.

Combining this criterion with \eqref{eq:sos} and \eqref{eq:aboveempty},
\begin{equation}
\label{eq:Scomplete}
S=\{(a,0):0<a<1\}\cup \{(a,\eta):0<a<1,\ \eta>0,\ F_{24}(a,\eta)\ge0\}.
\end{equation}

\section{BF violation at large radius and for anti-aligned charge}\label{sec:large}

\subsection{Large-radius limit and the level-four minimum}\label{sec:largeradius}

\begin{figure}[tbp]
\centering
\includegraphics[width=\textwidth]{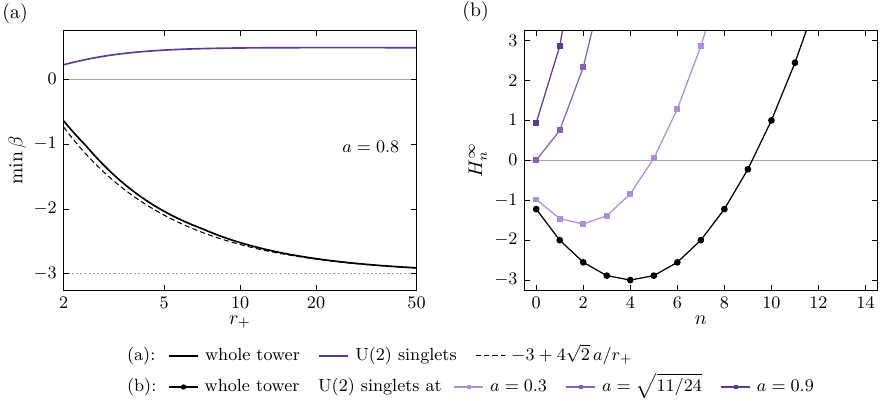}
\caption{BF discriminants at large horizon radius. (a) At fixed rotation $a=0.8$, the black curve is the minimum over the complete $t^+$ tower and the purple curve the minimum over its U(2) singlets. The horizontal axis $r_+$ is logarithmic. The dashed black curve gives $-3+4\sqrt2\,a/r_+$, and the dotted gray line marks the limiting value $-3$. The singlets pass while angular harmonics violate the bound. (b) The horizontal coordinate is the internal level $n$ and the vertical coordinate is the large-radius limit of each curve. Black circles give the level minimum $H_n^\infty$ over all horizon harmonics, which is negative for $n\le9$ and smallest at $n=4$. Purple squares give the singlet limits at $a=0.3$, $\sqrt{11/24}$ and $0.9$, from light to dark. The gray horizontal line is the BF threshold. Connecting lines guide the eye between integer levels. The minimizing spin grows as $j\propto r_+$ at fixed rotation.}
\label{fig:large}
\end{figure}

At fixed $0<a<1$ and $r_+\to\infty$, the minimum is attained at level four, with $j\propto r_+$ (figure~\ref{fig:large}). Here $s=\sqrt2\,r_+[1+O(r_+^{-2})]$.

The ellipse data of \eqref{eq:ellipse} behave as
\begin{equation}
\label{eq:largeS}
\begin{gathered}
\bmin=-3+\frac{8a}{s}+O(s^{-2}),\qquad \Qmin=6-\frac{6a}{s}+O(s^{-2}),\\
\frac{\Delta}{\ajj}\to\frac19,\qquad \ajj=\frac{8(1-a^2)^2}{3s^2}+O(s^{-3}),\qquad \jst(6)=\frac{3a\,s}{1-a^2}+O(1).
\end{gathered}
\end{equation}

Choose $j_\circ$ to be a half-integer nearest to $\jst(6)$. The level-four mode satisfies
\begin{equation}
\label{eq:levelfour}
\beta_{4,j_\circ,-j_\circ}\le\bmin+\frac{\Delta}{\ajj}(6-\Qmin)^2+\frac{\ajj}{16} =-3+\frac{8a}{s}+O(s^{-2}).
\end{equation}
All other levels have $\beta\ge-3+\frac19+O(s^{-1})$, and the tail $Q\ge12$ has $\beta\ge1-o(1)$ uniformly in $Q$ (appendix~\ref{app:large}). Consequently,
\begin{equation}
\label{eq:largemin}
\min_{n,j,k}\beta_{n,j,k}=-3+\frac{4\sqrt2\,a}{r_+}+O(r_+^{-2}).
\end{equation}
The estimates hold at fixed rotation and are not uniform at $a=0,1$. The minimizing mode is dual to a chiral primary of dimension six; its spin grows linearly with $r_+$, but its angular momentum per unit horizon length remains finite.

The competition between the levels is already visible before the minimization over the spin. Introduce the scaling variable $j=\varsigma s$ at fixed level. Then
\begin{equation}
\label{eq:scalinglimit}
\beta_{n,j,-j}\longrightarrow\frac{(Q-6)^2-27}{9}+\frac23\big(aQ-2(1-a^2)\varsigma\big)^2 .
\end{equation}
Minimizing over $\varsigma\ge0$ gives
\begin{equation}
H_n^\infty=\frac{(n^2-8n-11)}{9}.
\end{equation}
Hence, for sufficiently large $r_+$ at fixed $a\in(0,1)$, each level $n\le9$ contains a violating harmonic, whereas every harmonic at every level $n\ge10$ satisfies the bound. The uniform tail estimate above controls the latter statement over the infinite tower. For the singlets, $\varsigma=0$, the limit becomes
\begin{equation}
\frac19[(1+6a^2)Q^2-12Q+9].
\end{equation}
It is nonnegative for all $Q\ge2$ if and only if $a\ge\sqrt{11/24}$. At this threshold the next term in the expansion $(1320s-1511)/(1296s^2)$ is positive, so the singlets satisfy the bound there as well. Thus, all singlets satisfy the bound at large $r_+$ if and only if $a\ge\sqrt{11/24}$.

The minimizing spin $\varsigma=aQ/[2(1-a^2)]$ cancels the leading fiber momentum $k+QG$. The remaining value $((Q-6)^2-27)/9$ is the zero-momentum value of $\beta$ on the extremal planar Reissner--Nordstr\"om brane. Its zero set agrees with the previously obtained onset curve $R_{\mathcal O}^{\,2}=\frac23\Delta_{\mathcal O}(\Delta_{\mathcal O}-4)+2$~\mbox{\cite{Gubser:2009qm}}. With $\Delta_{\mathcal O}=Q$ and $R_{\mathcal O}=\frac23Q$, one has $\frac12[\frac23\Delta_{\mathcal O}(\Delta_{\mathcal O}-4)+2-R_{\mathcal O}^{\,2}]=((Q-6)^2-27)/9$. A singlet retains the fiber energy, whose level-zero contribution $\frac83a^2$ offsets $-\frac{11}9$ exactly at $a=\sqrt{11/24}$.

\subsection{BF violation with charge anti-aligned to rotation}\label{sec:antialigned}
The two extremal families with $q_C<0$ in section~\ref{sec:branch} lie outside $\Dom$. On the first, which has $\Omega_H>1$, the coefficient of $j^2$ in $\beta_{n,j,\pm j}$ is negative, independently of $n$, so every level is unbounded below as $j\to\infty$.

On the second family, which has $\Phi_H<-1$, continuing \eqref{eq:betaclosed} under $s\to-s$ gives the ellipse form \eqref{eq:ellipse}, now minimized at $k=+j$. Either the mode nearest the center or a singlet violates the bound at every black hole. The inequalities in appendix~\ref{app:antialigned} establish this throughout the family, although the required quantum numbers are not uniformly bounded at the parameter-space endpoints. Together, the two cases show that every extremal equal-rotation black hole with $q_C<0$ has a violating $t^+$ mode.

\section{Discussion}\label{sec:disc}

The singlet analysis leaves an above-BPS passing region~\mbox{\cite{Eloy:2026jzg}}. Including all horizon harmonics removes that region and confines the below-BPS passing set to a thin strip on the aligned domain. The mechanism is apparent for the minimizing harmonics $k=-j$: the angular energy on the base is linear in spin, since $j(j+1)-k^2=j$, while the fiber energy competes with the squared effective AdS$_2$ charge at quadratic order. Above the BPS curve this competition makes the discriminant unbounded below. Below it, the positive-definite quadratic part confines negative values to an ellipse, and the bound is satisfied when that ellipse contains no allowed charge--spin pair. The surviving region is thus a consequence of charge and spin discreteness.

The near-BPS and large-radius limits probe different physical scales. Approaching the BPS curve from above at a fixed rotation without a saturating mode requires both the internal level and the horizon spin to diverge. Every finite level cutoff consequently leaves a passing layer; at $a=1/5$ its width decreases as the inverse square of the cutoff. Interpreting this two-derivative result at finite string scale requires additional control. At large radius and fixed rotation, the minimizing internal level stays at four, while the spin grows proportionally to the radius. The angular momentum per unit horizon length remains finite, and the leading gauge-shifted fiber momentum cancels. The minimum then approaches the zero-momentum BF value of the extremal planar Reissner--Nordstr\"om brane~\mbox{\cite{Gubser:2009qm}}.

In the dual $\mathcal N=4$ super-Yang--Mills theory, level $n$ corresponds to a protected scalar primary of dimension $n+2$, while the horizon harmonic specifies the spatial angular dependence of its fluctuation on $S^3$~\mbox{\cite{Kim:1985ez,Gunaydin:1984fk,Witten:1998qj}}. The large-radius minimum probes a dimension-six primary; above-BPS violation near a nonsaturating BPS rotation instead probes increasing operator dimension and spatial angular momentum. The lowest BF discriminant at zero temperature need not identify the mode with the highest onset temperature~\mbox{\cite{Gubser:2009qm}}. Boundary retarded response depends on matching the throat solution to the asymptotic AdS region~\mbox{\cite{Faulkner:2009wj}}.

The threshold at unit horizon angular velocity is also familiar from rotational superradiance~\mbox{\cite{Cardoso:2004hs,Berkooz:2013tg,Green:2015kur}}. The Hawking--Reall energy argument requires a globally timelike horizon Killing field, matter obeying the dominant energy condition and no energy influx at the AdS boundary~\mbox{\cite{Hawking:1999dp}}. Charged perturbations also carry an electrostatic contribution to their conserved energy~\mbox{\cite{Green:2015kur}}. Neither the causal character of the five-dimensional Killing field nor the uplift argument by itself establishes positive perturbation energy for the $t^+$ sector, whose fields combine metric and four-form fluctuations.

Finite-temperature singlet onsets and hairy black-hole branches have been found for this family~\mbox{\cite{Dias:2024edd}}. They do not settle stability at exactly zero temperature, where a growth rate may vanish or the limiting boundary-value problem may become singular. Small Reissner--Nordstr\"om--AdS$_5$ black holes illustrate the limitation of the local criterion: sufficiently near extremality, a charge-two scalar of mass squared $-4$ is superradiantly unstable~\mbox{\cite{Bhattacharyya:2010yg}}. Its exactly extremal level-zero singlet nevertheless obeys the BF bound at sufficiently small radius.

Other supergravity and Kaluza--Klein sectors may reduce the passing set~\mbox{\cite{Ezroura:2024xba,Eloy:2026jzg}}. Their intersection with the set found here remains inside the outer strip, but need not retain its accumulation along the BPS curve. Localization on $S^5$ supplies a distinct class of perturbations~\mbox{\cite{Hubeny:2002xn,Murata:2008xr,Dias:2015pda}}. Dual-giant emission at electric potential greater than one is another proposed decay process below the BPS curve, involving D3-branes rather than linear Kaluza--Klein fluctuations~\mbox{\cite{Choi:2024xnv}}. Hairy black holes, grey galaxies and dual-dressed configurations provide possible nonlinear phases~\mbox{\cite{Dias:2024edd,Kim:2023sig,Bajaj:2024utv,Choi:2024xnv}}. Identifying any of them as an endpoint of the modes studied here requires additional dynamical evidence.

The level-four mode at large radius provides a concrete target for a radial analysis, since its internal level remains finite. Solving its radial equation would determine whether the throat BF violation corresponds to a growing mode satisfying the asymptotic AdS boundary condition. The large-spin modes above the BPS curve and the singlets in the BF-passing strip give complementary radial problems. Extending the harmonic analysis to other towers, unequal rotations or independent electric charges would test how broadly the charge--spin mechanism applies.

\acknowledgments

We would like to thank Timm Wrase, Sera Cremonini and Finn Larsen for discussions. The work of Y.T. is supported by the National Natural Science Foundation of China Grants No. 12247103. The work of X.L. and Z.M. is supported in part by the NSF grant PHY-2609419. Z.M. acknowledges the support of the Dr. Hyo Sang Lee Graduate Fellowship from the College of Arts and Sciences at Lehigh University. 

\paragraph*{AI assistance.}
The authors used several large language models (Codex Astra, Claude Opus) to assist with writing, editing, and checking calculations in this paper. The authors have carefully reviewed all such contributions and take full responsibility for the content, accuracy, and conclusions of the work.

\appendix
\section{The scalar equation and near-horizon reduction}\label{app:A}

We derive the scalar equation used in section~\ref{sec:setup}, including the synchronization condition and the vanishing Pauli coupling that enters the BPS analysis. We retain the CCLP coordinates $(t_C,r,\theta_C,\phi_C,\psi_C)$ and gauge field $\AF$ in the convention of footnote~\ref{fn:conv}, with horizon potential $\Phi_H$ given by \eqref{eq:OmPhi}. We write $\mathcal F=d\AF$, $X=r^2$ and $X_+=r_+^2$, and use $s=\sqrt{1+2a^2+2r_+^2}$, $\eta=s-1-2a$, $\Theta_1=3s^2-1-2a^2$, $\Theta_2=s-1+2a^2$ and $U=s^2-s-2a^2$ as in the main text.

\subsection{Extremal backgrounds, regularity and near-horizon limit}\label{app:nhdata}

We first give the full radial equation and identify the two regular extremal charge roots. The subsequent near-horizon limit fixes the metric coefficients and the synchronization frequency.

In the nonrotating Euler angles $(\theta,\phi,\psi^0)$ of \eqref{eq:angles}, with left-invariant forms $\sigma_i^0$,
\begin{equation}
\sigma_3^0=d\psi^0+\cos\theta\,d\phi =2(\sin^2\theta_C\,d\phi_C+\cos^2\theta_C\,d\psi_C),\qquad \AF=\frac{q_C(dt_C-\frac a2\sigma_3^0)}{(1-a^2)(X+a^2)}.
\end{equation}
The charged wave equation separates for arbitrary CCLP parameters $(q_C,m_C)$~\mbox{\cite{Wu:2009separation}}. For charge $Q$ and mass squared $M^2$, the mode $e^{-i\omega t_C}R(X)D^j_{mk}(\theta,\phi,\psi^0)$ uses the Wigner functions of section~\ref{sec:nh} and obeys $(\Hor R')'+V_{\rm eff}R=0$ in this gauge, with $'=d/dX$. Here $\Hor$ is $r^2$ times the radial metric function, and
\begin{equation}
\label{eq:Veff}
\begin{aligned}
\Hor&=(X+a^2)^2(1+X)+q_C^2+2a^2q_C-2m_CX,\\
\Upsilon&=(1-a^2)(X+a^2)^3+2a^2(m_C+q_C)(X+a^2)-a^2q_C^2,\\
V_{\rm eff}&=\frac{\Upsilon\,(\tilde\omega-\Omega_3W_3)^2}{4\Hor}-\frac{(1-a^2)^2(X+a^2)^3\,W_3^2}{\Upsilon}\\
&\quad-(1-a^2)\big[j(j+1)-k^2\big]-\frac{M^2(X+a^2)}{4},
\end{aligned}
\end{equation}
where
\begin{equation}
\label{eq:Veff2}
\begin{gathered}
\tilde\omega=\omega+\frac{Q\,q_C}{(1-a^2)(X+a^2)},\qquad W_3=k+\frac{a\,Q\,q_C}{2(1-a^2)(X+a^2)},\\
\Omega_3=\frac{2a\big[(X+a^2)\big(2m_C+(1+a^2)q_C\big)-q_C^2\big]}{\Upsilon}.
\end{gathered}
\end{equation}

A degenerate horizon at $X_+$ requires $\Hor'(X_+)=0$, fixing $m_C$ as in \eqref{eq:qm}. The remaining condition $\Hor(X_+)=0$ is $(q_C+a^2)^2=s^2r_+^4$. Its roots are $q_C=s\,r_+^2-a^2$, as in \eqref{eq:qm}, and $q_C=-a^2-s\,r_+^2$, obtained by $s\to-s$. Both give $\Hor=(X-r_+^2)^2(X+s^2)$, so $r_+$ is the only horizon, as in \eqref{eq:Hor}.

To check regularity, use the frame $(dt_C,dX,\sigma_i^0)$. The Hopf base has norm $(X+a^2)/(4(1-a^2))$, the $(t_C,\sigma_3^0)$ block has determinant $-\Hor/(4(1-a^2)^2(X+a^2))$, and the fiber has norm $\Upsilon/(4(1-a^2)^2(X+a^2)^2)$. Expanding in $Y=X-r_+^2$ gives $\Upsilon=\sum_{i=0}^3\Upsilon_iY^i$, with $\Upsilon_2=\frac32(1-a^2)(s^2-1)$ and $\Upsilon_3=1-a^2$ on both roots. The remaining coefficients on the first and second roots are, respectively,
\begin{equation}
\label{eq:fibrenorm}
\begin{aligned}
\Upsilon_0&=\tfrac14(s+1)^2\Theta_2^2\,r_+^2, &\qquad \Upsilon_0&=\tfrac14(s-1)^2(s+1-2a^2)^2\,r_+^2,\\
\Upsilon_1&=\tfrac14(s+1)^2\Theta_2\,(3s-3-2a^2), &\qquad \Upsilon_1&=\tfrac14(s-1)^2(s+1-2a^2)(3s+3+2a^2).
\end{aligned}
\end{equation}
All $\Upsilon_i$ are positive for $0<a<1$ and $r_+>0$. For the first-root coefficient $\Upsilon_1$, this follows from $s>\sqrt{1+2a^2}\ge1+(\sqrt3-1)a^2$, which gives $3s-3-2a^2>(3\sqrt3-5)a^2$. Thus the angular metric is positive definite on and outside the horizon. Outside it, the negative block determinant gives $g^{t_Ct_C}<0$, making $t_C$ a time function and excluding closed causal curves. The positive horizon fiber norm also meets the regularity criterion~\mbox{\cite{Chong:2005hr}}. Both roots therefore describe regular black holes. On the first root, $q_C\ge0$ requires $s\ge\frac12\big(1+\sqrt{1+8a^2}\big)$, equivalently $2r_+^4+r_+^2\ge a^2$. On the second, $q_C<-a^2$.

\phantomsection\label{app:closedform}
After $\psi^0=\psi+2\Omega_Ht_C$, with $\Omega_H$ of \eqref{eq:OmPhi}, the horizon generator is $\partial_{t_C}$ at fixed $\psi$ and $(\AF)_{t_C}=\Phi_H$ at the horizon; the gauge change $\AF\to\AF-\Phi_Hdt_C$ multiplies a charge-$Q$ scalar by $e^{-iQ\Phi_Ht_C}$, so
\begin{equation}
\label{eq:modechange}
e^{-i\omega t_C}D^j_{mk}(\theta,\phi,\psi^0) \longmapsto e^{-i(\omega-2k\Omega_H+Q\Phi_H)t_C}D^j_{mk}(\theta,\phi,\psi).
\end{equation}
The scaling $r-r_+=\epsilon x$, $t_C=t/\epsilon$ gives the throat frequency $\hat\omega=(\omega-\omega_s)/\epsilon$, with $\omega_s=2k\Omega_H-Q\Phi_H$. The limit is \eqref{eq:nhmetric}, with
\begin{equation}
\label{eq:nhconstants}
\begin{aligned}
A^2&=\frac{4(s-1)\Theta_1}{(1+s)\Theta_2^2}, & B^2&=\frac{s^2-1}{4\Theta_1},\\
C^2&=\frac{s^2-1}{2(1-a^2)}, & D^2&=\frac{\Theta_2^2\,(s^2-1-2a^2)}{2(1-a^2)^2(s-1)^2},\\
E&=\frac{8\sqrt2\,a(1-a^2)(2s^2-s-1-2a^2)}{(1+s)\Theta_2^2\sqrt{s^2-1-2a^2}}, & F&=\frac{2\sqrt2\,\sqrt{s^2-1-2a^2}\;U}{(s^2-1)\Theta_2},\\
G&=\frac{a\,U}{2(1-a^2)(s-1)},
\end{aligned}
\end{equation}
where only $C^2$, given in \eqref{eq:C2} and discussed in footnote~\ref{fn:C2}, differs from the previously reported expressions~\mbox{\cite{Eloy:2026jzg}}. With $\ell_2=B$, \eqref{eq:nhmetric} and \eqref{eq:nhconstants} give
\begin{equation}
\label{eq:Fsquared}
\mathcal F^2=\frac{32G^2}{C^4}-\frac{2F^2}{A^2B^2},\qquad \ell_2^2\mathcal F^2\big|_{\eta=0}=-\frac{2}{1+5a}.
\end{equation}
For $\xi_i$ dual to the one-forms $\sigma_i$ in \eqref{eq:nhmetric}, the squashed-sphere Laplacian $4(\xi_1^2+\xi_2^2)/C^2+4\xi_3^2/D^2$ has eigenvalue $-4[j(j+1)-k^2]/C^2-4k^2/D^2$ on $D^j_{mk}$. The magnetic potential shifts $k\to k+QG$, while the electric potential and fiber twist give $AB\,q_2=QF-E(k+QG)$. These expressions yield \eqref{eq:m2q2}; substituting the constants \eqref{eq:nhconstants} into \eqref{eq:beta} then gives \eqref{eq:betaclosed}.

\subsection{Horizon exponents of the full radial equation at synchronization}\label{app:horexp}

At synchronization, the full radial equation has a regular singular point at the extremal horizon. Its indicial equation recovers the same BF discriminant as the throat reduction.

By \eqref{eq:Hor}, $\frac12\Hor''(X_+)=3X_++1+2a^2=\Theta_1/2$; also $\Omega_3(X_+)=2\Omega_H$. The first term of $V_{\rm eff}$ in \eqref{eq:Veff} thus has a double pole at $X_+$ unless $\tilde\omega=\Omega_3W_3$ there, i.e.\ unless $\omega=\omega_s$. At $\omega=\omega_s$, $V_{\rm eff}$ has a finite horizon limit $V_0$, and near $X_+$ the equation becomes $\frac{\Theta_1}2[(X-X_+)^2R']'+V_0R=0$, solved by $(X-X_+)^{\sigma}$ with
\begin{equation}
\label{eq:frob}
\sigma^2+\sigma+\frac{2V_0}{\Theta_1}=0,\qquad \sigma=-\frac12\pm\frac12\sqrt{1-\frac{8V_0}{\Theta_1}} ,
\end{equation}
and evaluating $V_0$ gives $1-8V_0/\Theta_1=\beta$ identically in $(M^2,Q,j,k)$, which is \eqref{eq:horexp}. Through its coefficient of $j(j+1)$, this identity confirms \eqref{eq:C2} independently. Since \eqref{eq:Veff} holds for all $(q_C,m_C)$ and both charge roots give \eqref{eq:Hor}, the identity holds on both, with $\beta$ given by \eqref{eq:betaclosed} on the first and at $s\to-s$ on the second.

\subsection{Decoupling and vanishing Pauli coupling of the \texorpdfstring{$t^+$}{t+} tower}\label{app:pauli}

We verify the decoupling, absence of mixing and $\fp_n=0$ in \eqref{eq:pauliconvention}, and compare the level-zero result with the nonzero coupling reported by Ezroura and Larsen~\mbox{\cite{Ezroura:2024xba}}.

At level zero, we use the relevant terms in the $\mathrm{SL}(6)/\mathrm{SO}(6)$ sector of $\mathcal N=8$, SO(6) gauged supergravity~\mbox{\cite{Cvetic:2000nc}}:
\begin{equation}
\label{eq:Lsl6}
\begin{aligned}
\mathcal L={}&R{*1}-\tfrac14\mathcal T^{-1}_{ij}{*D\mathcal T_{jk}}\wedge\mathcal T^{-1}_{kl}D\mathcal T_{li} -\tfrac14\mathcal T^{-1}_{ik}\mathcal T^{-1}_{jl}{*F^{ij}}\wedge F^{kl}-V{*1},\\
V={}&\tfrac12\big(2\mathcal T_{ij}\mathcal T_{ij}-(\mathcal T_{ii})^2\big),
\end{aligned}
\end{equation}
with unit gauge coupling, symmetric unimodular $\mathcal T$, SO(6) gauge field $A^{ij}$, field strength $F^{ij}$, and $D\mathcal T=d\mathcal T+[A,\mathcal T]$. The minimal theory, with Einstein--Maxwell part $R+12-\frac34\mathcal F^2$, is embedded by $\mathcal T=\mathbf1$ and $A^{ij}=\AF\mathbb J_{ij}$, where $\mathbb J^2=-\mathbf1$. Writing $\mathcal T=e^\varphi$, the symmetric traceless $\varphi\in\mathbf{20}'$ splits into the twelve-dimensional $t^+$, anticommuting with $\mathbb J$, and $t^-$, commuting with it. On $t^+$, $\operatorname{ad}_{\mathbb J}\varphi\equiv[\mathbb J,\varphi]=2\mathbb J\varphi$ and $\operatorname{ad}_{\mathbb J}^2=-4$, giving $|Q|=2$; $t^-$ is neutral.

The identity $e^{-\varphi}\mathbb J=\mathbb Je^{\varphi}$ on $t^+$ gives $\mathcal T^{-1}\mathbb J\mathcal T^{-1}=\mathbb J$, so its gauge kinetic function $f_{\rm kin}=\operatorname{Tr}(\mathbb J^T\mathcal T^{-1}\mathbb J\mathcal T^{-1})$ equals six to all orders. On $t^-$, instead, $f_{\rm kin}=6+2\operatorname{Tr}\varphi^2+\cdots$. In both sectors $V=-12-\operatorname{Tr}\varphi^2+O(\varphi^3)$.

To fix the normalization, take $\varphi=\sum_Ac_AE_A$ with $\operatorname{Tr}(E_AE_B)=2\delta_{AB}$. Terms $V\supset\kappa_V\operatorname{Tr}\varphi^2$ and $f_{\rm kin}\supset\kappa'\operatorname{Tr}\varphi^2$ then give $-\frac12(Dc_A)^2-2\kappa_Vc_A^2-\frac14\kappa'c_A^2\mathcal F^2$. Thus \eqref{eq:pauliconvention} has $M^2=4\kappa_V=-4$ and $\fp=-\kappa'$, equal to $0$ on $t^+$ and $-2$ on $t^-$.

The twelve $t^+$ scalars decouple at quadratic order. Orthogonality of $t^\pm$ under the trace is preserved by $\operatorname{ad}_{\mathbb J}$ and excludes kinetic cross terms; $V$ and $f_{\rm kin}$ have none either. For $f_{\mu\nu}^{ij}=\delta F_{\mu\nu}^{ij}$, the scalar--vector bilinear is proportional to $-\mathcal F_{\mu\nu}\operatorname{Tr}[(\varphi\mathbb J+\mathbb J\varphi)f^{\mu\nu}]$ and vanishes on $t^+$. The remaining scalars outside the $\mathbf{20}'$ and the twelve propagating two-forms carry charge under the unbroken $\mathrm{SO}(2)\subset\mathrm{SL}(2,\mathbb R)$, unlike $t^+$, so they cannot mix with it~\mbox{\cite{Ezroura:2024xba}}.

There is no linear $t^+$ term in the stress tensor: $V$ and $f_{\rm kin}$ have no linear term, and the scalar kinetic term starts at quadratic order since $D\mathcal T=0$ on the background. The Chern--Simons term contains neither $\mathcal T$ nor the metric, so metric perturbations also decouple. The consistent type IIB embedding lifts these level-zero perturbations without sourcing discarded fields~\mbox{\cite{Cvetic:2000nc,Baguet:2015sma}}.

Ezroura and Larsen define $t^\pm$ by $\Omega t^\pm\Omega=\pm t^\pm$~\mbox{\cite{Ezroura:2024xba}}. This is our split for $\Omega=\mathbb J$; their minimal-theory normalization and effective-mass convention also agree with ours. With $\mathcal T=\mathbf1+t+O(t^2)$, the action gives, before vector fluctuations are eliminated,
\begin{equation}
\label{eq:paulilinearization}
D^\mu D_\mu t+4t+\tfrac12(\Omega t\Omega-t)\mathcal F^2=(\text{vector terms}).
\end{equation}
Their exact scalar equation contains the gauge-source matrix $-\mathcal T^{-1}\Omega\mathcal T^{-1}\Omega$~\mbox{\cite{Ezroura:2024xba}}. The matrix expands as
\begin{equation*}
-\mathcal T^{-1}\Omega\mathcal T^{-1}\Omega=\mathbf1+\Omega t\Omega-t+O(t^2).
\end{equation*}
The two inverses contribute $-t$ and $\Omega t\Omega$, which cancel on $t^+$ and add to $-2t$ on $t^-$. With our conventions, replacing $\Omega t\Omega-t$ by $2\Omega t\Omega$ in \eqref{eq:paulilinearization} reproduces the reported linearized field-strength terms in the charged and neutral sectors~\mbox{\cite{Ezroura:2024xba}}. This agrees with the action for the neutral sector, with $\fp=-2$ before vector fluctuations are eliminated, but gives $\fp_0=2$ for $t^+$. The same charged value enters their near-horizon effective mass. The discrepancy is therefore in the linearized scalar coupling, not in the effective-mass convention. The vanishing coupling also agrees with the charge-two, mass-squared $-4$ scalar in a consistent truncation whose Maxwell kinetic term is independent of that scalar~\mbox{\cite{Bhattacharyya:2010yg}}.

\phantomsection\label{app:alllevels}
At higher levels, the exceptional-field-theory expansion expresses degree-$n$ fluctuations as level-zero internal representations tensored with degree-$n$ scalar harmonics on $S^5$~\mbox{\cite{Malek:2020yue}}. Internal derivatives act as SO(6) generators, independently of the five-dimensional background, so the covariant linearized equations preserve $n$~\mbox{\cite{Eloy:2026jzg}}. The identity background scalar matrix and vector proportional to $\mathbb J$ also preserve $\mathrm{SU}(3)\times\mathrm{U}(1)\times\mathrm{SO}(2)$, where SO(2) stabilizes the constant axio-dilaton. Mixing therefore requires the same representation at the same harmonic degree.

With SU(3) representations denoted by $[p,q]$ and charges by subscripts, the SO(2)-neutral degree-$n$ harmonics branch as $\bigoplus_{i=0}^n[i,n-i]_{2i-n}$; only $[n,0]$ reaches charge $n$. Among level-zero scalars, only the $\mathbf{20}'$ is SO(2) neutral, with top charge two in $[2,0]$. An SO(2)-neutral scalar with $Q=n+2$ must therefore lie in
\begin{equation}
\label{eq:scalarrep}
[2,0]\otimes[n,0]=[n+2,0]\oplus[n,1]\oplus[n-2,2],
\end{equation}
where the last two summands require $n\ge1$ and $n\ge2$, respectively, and the $t^+$ representation $[n+2,0]$ occurs once. The SO(2)-neutral vector and two-form indices lie in the $\mathbf{15}$, with charge-two part $[0,1]$, giving the charge-$(n+2)$ representations
\begin{equation}
\label{eq:vectorrep}
[0,1]\otimes[n,0]=[n,1]\oplus[n-1,0],
\end{equation}
where the second summand requires $n\ge1$. Neither representation is $[n+2,0]$, and metric fluctuations reach only charge $n$. In the scalar gauge transformation, the external parameter multiplies the vanishing derivative of the constant background scalar matrix, while the internal parameter carries a vector index~\mbox{\cite{Eloy:2026jzg}}. Neither can shift the $[n+2,0]$ scalar. It is therefore a physical mode with no mixing partner at fixed degree.

For this scalar, with vanishing background scalar current and no mixing fields, the general scalar equation has only kinetic, mass and Pauli terms~\mbox{\cite{Eloy:2026jzg}}. The Pauli matrix is built from the background vector and constant scalar-index tensors without internal derivatives~\mbox{\cite{Eloy:2026jzg}}. It therefore acts as $\mathbb P\otimes\mathbf1$ on harmonic coefficients, with $\mathbb P$ the level-zero Pauli matrix. On the maximal-charge subspace the scalar factor lies in the charge-two $[2,0]$ of $t^+$, where symmetry makes $\mathbb P$ a multiple of the identity, zero by the level-zero result. Projecting to $[n+2,0]$ gives \eqref{eq:pauliconvention} with $\fp_n=0$ and $M^2=n^2-4$.

\section{The quadratic form and fixed-level BF boundaries}\label{app:B}

The fixed-level analysis uses the quadratic coefficients, their signs and the local expansion at BPS saturation. Throughout this appendix the black holes lie in $\Dom$. We retain $s=\sqrt{1+2a^2+2r_+^2}$, the positive denominators $\Theta_1=3s^2-1-2a^2$ and $\Theta_2=s-1+2a^2$, and $U=s^2-s-2a^2=2q_C/(1+s)\ge0$. The BPS curve is $\eta=s-1-2a=0$.

\subsection{Fiber momentum minimization and quadratic-form coefficients}\label{app:kmin}

At fixed $(n,j)$, \eqref{eq:betaclosed} is quadratic in $k$ with
\begin{equation}
\label{eq:kcurv}
\partial_k^2\beta_{n,j,k}=-\frac{48a^2(1-a^2)(s^2-1)(1+s)^2}{\Theta_1^2\Theta_2^2}\le0 ,
\end{equation}
so its minimum over $|k|\le j$ lies at $k=\pm j$, and, by \eqref{eq:kdiff}, at $k=-j$, uniquely if $j>0$ and $U>0$.

\phantomsection\label{app:coeffs}%
Setting $k=-j$ and $M^2=Q^2-4Q$ in \eqref{eq:betaclosed} gives \eqref{eq:form} and \eqref{eq:coeffs}, with
\begin{equation}
\label{eq:NQQ}
\begin{aligned}
\NQQ={}&1-2a^2+4a^4+16a^6-4s+12a^2s+8a^4s+6s^2-12a^2s^2\\
&-12a^4s^2-4s^3-4a^2s^3+s^4+6a^2s^4 .
\end{aligned}
\end{equation}
In terms of $\eta$, the factors
\begin{equation}
\label{eq:etaforms}
\begin{gathered}
\Theta_1=2+12a+10a^2+6\eta+12a\eta+3\eta^2,\qquad \Theta_2=2a+2a^2+\eta,\\ 1+s=2+2a+\eta,\qquad s-1+2a=4a+\eta
\end{gathered}
\end{equation}
and
\begin{equation}
\label{eq:Njjeta}
\Njj=8a^2(1+a)^2+8a(1+a)(1+6a)\,\eta+2(1+18a+30a^2)\,\eta^2+6(1+4a)\,\eta^3+3\eta^4 ,
\end{equation}
all have positive coefficients for $a>0$ and are therefore positive for $\eta\ge0$.

To prove \eqref{eq:aQQpos}, put $w=s-1$ and $a^2=\chi w(w+1)/2$; then $U=w(w+1)(1-\chi)$, so $0<\chi\le1$ covers $\Dom$, and
\begin{equation}
\label{eq:NQQwchi}
\begin{aligned}
\NQQ={}&2\chi(3-2\chi+\chi^2)\,w^3+(1+16\chi-11\chi^2+6\chi^3)\,w^4\\
&+\chi(13-10\chi+6\chi^2)\,w^5+\chi(3-3\chi+2\chi^2)\,w^6,
\end{aligned}
\end{equation}
with positive brackets: $3-2\chi+\chi^2=2+(1-\chi)^2$, $16\chi-11\chi^2\ge0$, and $13-10\chi+6\chi^2$ and $3-3\chi+2\chi^2$ have negative discriminants. Hence $\NQQ>0$ and $\aQQ>0$.

The determinant \eqref{eq:Delta} follows from \eqref{eq:coeffs} and \eqref{eq:OmPhi}. With $c(a)$ of \eqref{eq:zeta}, \eqref{eq:bpsD} gives on the BPS curve
\begin{equation}
\label{eq:bpscoeffs}
\begin{gathered}
\aQQ=\frac{16a^2}{(1+5a)^2},\qquad \aQj=-\frac{8a(1-a)}{(1+5a)^2},\qquad \ajj=4c(a)^2,\\
\aQ=-\frac{8a}{1+5a},\qquad \aj=\frac{4(1-a)}{1+5a}.
\end{gathered}
\end{equation}
These give \eqref{eq:perfectsquare} and, for the vertex $y_n$ of \eqref{eq:spinsquare}, $y_n=\zeta_n$ of \eqref{eq:zeta}.

\phantomsection\label{app:ergo}%
For the sign of $\ajj$ in sections~\ref{sec:largelabels} and~\ref{sec:above}, let $t$, $x$ be the AdS$_2$ coordinates of \eqref{eq:nhmetric}; the constants \eqref{eq:nhconstants} give
\begin{equation}
\label{eq:ergo}
g(\partial_t,\partial_t)=x^2\left(\frac{D^2E^2}{4}-A^2\right),\qquad \frac{D^2E^2}{4}-A^2=-\frac{8r_+^2\Njj}{(s^2-1)^2\Theta_2^2}.
\end{equation}
As $\ajj$ is a positive multiple of $\Njj$, $\ajj<0$ exactly where the horizon generator $\partial_t$ is spacelike in the throat. The numerator $\Njj=2U^2+2(s^2-1)U-(s^2-1)^2$ increases with $U\ge0$ and vanishes at $2U=(\sqrt3-1)(s^2-1)$. At fixed $a$, the ratio $2U/(s^2-1)$ has $s$-derivative $2[(s-1)^2+4a^2s]/(s^2-1)^2>0$ and increases from zero at $q_C=0$ to one on the BPS curve, so the region $\ajj<0$ is
\begin{equation}
\label{eq:ergoregion}
\tfrac12\big(1+\sqrt{1+8a^2}\big)\le s<s_0(a),\qquad s_0(a)=\frac{1+\sqrt{(2-\sqrt3)^2+4(3-\sqrt3)a^2}}{3-\sqrt3},
\end{equation}
above the BPS curve.

For the minimizing harmonic of section~\ref{sec:lowestmode}, let $q_C>0$, so $U>0$ and $\aQj<0$. By \eqref{eq:quadj}, $\beta_{n,\frac12,-\frac12}-\beta_{n,0,0}=\ajj/4+\lambda_n/2$ has slope $\aQj$ in $Q$ and is negative at large $n$. Where $\ajj>0$, the minimizing spin $\hj_n$ of \eqref{eq:levelmin} has $|2\hj_n-y_n|\le\frac12$ once $y_n\ge0$, and $\partial_Qy_n=-2\aQj/\ajj$, so $\hj_n/n\to-\aQj/\ajj$.

\subsection{Saturation, onset of BF violation and minimization over real spin}\label{app:sos}

Substituting \eqref{eq:bpsD} into \eqref{eq:betaclosed} gives \eqref{eq:sos}, which at $k=-j$ reduces to \eqref{eq:alongcurve}. For $k>-j$, $(j+k)(j-k+1)\ge2j\ge1$, so $\beta^{\rm BPS}\ge4(1-a)/(1+5a)>0$.

The remainder of \eqref{eq:levelsquare} is $\eta\,\GG$ with
\begin{equation}
\label{eq:Gcoeffs}
\GG=\frac{\Delta}{\eta\,\aQQ}\,j^2+\frac1\eta\Big(\aj-\frac{\aQj\aQ}{\aQQ}\Big)j+\frac1\eta\Big(1-\frac{\aQ^2}{4\aQQ}\Big),
\end{equation}
whose numerators vanish on the BPS curve by \eqref{eq:alongcurve}. The common denominator, $\Theta_1\NQQ$, is positive (appendix~\ref{app:kmin}). Thus $\GG$ is regular at $\eta=0$, with leading coefficient
\begin{equation}
\label{eq:G2}
\frac{\Delta}{\eta\,\aQQ}=\frac{8(1-a)^2(1+a)^2(2a+\eta)^2(4a+\eta)}{\Theta_1\,\NQQ}>0 ;
\end{equation}
and on the BPS curve, with $v_j$ of \eqref{eq:levelsquare},
\begin{equation}
\label{eq:Gbps}
\begin{gathered}
v_j=\frac{(1+2j)-(3+2j)\,a}{4a},\\ \GG(j;a,0)=\frac{4(1-a)^2j^2-4(1-a)(1+7a)\,j+(5a-3)(1+5a)}{4a(1+a)(1+5a)}.
\end{gathered}
\end{equation}
At $a=a_c$ of \eqref{eq:ac}, $v_j=n$, so \eqref{eq:levelsquare} gives $\partial_\eta\beta=\GG(j;a_c,0)$, and analyticity with \eqref{eq:alongcurve} yields \eqref{eq:tangexp}. Evaluating \eqref{eq:Gbps} at $a=a_c$ gives \eqref{eq:P}, and substituting $a_c$ into $\mathcal P(a,n+2)$ gives \eqref{eq:Pa}.

\phantomsection\label{app:realspin}%
By \eqref{eq:coeffs}, the minimum $\mu_n$ over real spin in \eqref{eq:spinsquare} factors as
\begin{equation}
\label{eq:mufact}
\begin{aligned}
\mu_n&=\eta\,\frac{8(1-a^2)^2(1+s)}{\ajj\,\Theta_1^3\Theta_2^2}\,\varpi(Q;a,s),\\
\varpi&=(s-1)^2(s+1)(s-1+2a)\,Q^2-4W Q+3(s-1)(s-1+2a)\Theta_1,
\end{aligned}
\end{equation}
with
\begin{equation}
\label{eq:W}
W=3s^4+3as^3-6s^3-6a^2s^2-4as^2+2s^2-2a^3s+6a^2s-as+2s+4a^4+2a^3+2a-1 .
\end{equation}
Apart from $\eta$ and $1/\ajj$, the prefactor is positive on $\Dom$, so where $\ajj>0$, $\mu_n<0$ if and only if $\eta\,\varpi(Q)<0$. On the BPS curve, \eqref{eq:bpsD} reduces $\varpi$ to $16a^2(1+a)\,\mathcal P(a,Q)$ and the prefactor divided by $\eta$ to $1/[8a^2(1+a)^2(1+5a)]$, which gives
\begin{equation}
\label{eq:Pderiv}
\partial_\eta\mu_n\big|_{\eta=0}=\frac{2\,\mathcal P(a,Q)}{(1+a)(1+5a)} ,
\end{equation}
with $Q^2$ coefficient $\tauh(a)$ of \eqref{eq:tauhat}.

For $n\ge3$, the linear function $\mathcal P(a,n+2)$ increases with $a$ and vanishes at $a_*(n)$ of \eqref{eq:astar}, where the identity $(2n+1)(2n-5)=2(2n^2-4n-1)-3$ implies $a_*(n)<1/(n-5/2)$. At low levels $n=0,1,2,3$ one finds $P(n,j)=-2-2j,\,-12-6j,\,-28-6j,\,-50-2j$, whereas at fixed low spins $j=0,1/2,1,3/2$ one obtains $P(n,j)=-(3n^2+7n+2),\,-(2n^2+10n+3),\,-(n^2+13n+4),\,-(16n+5)$, respectively.

Once $n\ge4$, the prefactor $n^2-3n-1$ in \eqref{eq:Tn} is positive while $T_n$ decreases strictly from $T_4=26$, because $(16n+5)/(n^2-3n-1)$ has derivative $-(16n^2+10n+1)/(n^2-3n-1)^2<0$; since $T_{20}=1342/339<4$, only spins $j\le3/2$ can satisfy $P<0$ for $n\ge20$.

Because $P<0$ for $n\le3$, any zero of $P$ requires $n\ge4$ and hence $(n^2-3n-1)\mid(16n+5)$ by \eqref{eq:Tn}; combining this with $256(n^2-3n-1)=(16n+5)^2-58(16n+5)+9$ forces $(n^2-3n-1)\mid9$, which leaves $n\in\{4,5\}$ and rules out $n=5$ via $9\nmid85$, so $P$ vanishes uniquely at $(n,j)=(4,13)$ where $2j=T_4=26$ and $a_c(4,13)=3/5$. With $u=a-3/5$,
\begin{equation}
\label{eq:degenerate}
\beta_{4,13,-13} =\frac{2025}{16}\,u^2-\frac{2265}{64}\,u\eta-\frac{3751}{1024}\,\eta^2 +O\big((|u|+|\eta|)^3\big).
\end{equation}
As $\frac{2025}{16}\cdot\big(-\frac{3751}{1024}\big)-\frac14\big(\frac{2265}{64}\big)^2=-\frac{795375}{1024}<0$, the saturation point is a nondegenerate saddle: by the Morse lemma, two analytic zero curves cross transversally there, and $\beta_{4,13,-13}<0$ in the two opposite sectors between them containing the line $a=\frac35$, where the $\eta^2$ coefficient is negative.

\subsection{BF boundaries near saturation and at large level}\label{app:neartangency}

At $a_c=a_c(n,j)$, where $\zeta_n=2j$, every other endpoint harmonic $(j',-j')$ of level $n$ has $\beta^{\rm BPS}\ge c(a_c)^2$ by \eqref{eq:alongcurve}. Modes with $k>-j'$ obey $\beta^{\rm BPS}\ge4(1-a)/(1+5a)$ by appendix~\ref{app:sos}, and $\ajj>0$ keeps large spins uniformly away from zero nearby. Thus $(j,-j)$ uniquely minimizes the level near $a_c$, with $H_n=\beta_{n,j,-j}$. For $P(n,j)\ne0$, the implicit function theorem gives a zero set tangent to the BPS curve; \eqref{eq:epseta} then converts \eqref{eq:tangexp} into \eqref{eq:tangphys}. The exceptional mode $(4,13)$ instead has the saddle expansion \eqref{eq:degenerate}.

\phantomsection\label{app:islands}
For $n=19,20$, the boxes
\begin{equation}
\begin{aligned}
B_{19}&=(0.06003,0.06036)\times(0,1.87\times10^{-3}),\\
B_{20}&=(0.05724,0.05789)\times(-3.59\times10^{-3},0)
\end{aligned}
\end{equation}
in $(a,\eta)$ have BPS edges containing $a_c(n,2)$. On the closed rectangles with the same $a$ endpoints and $\eta$ intervals $[0,19/10000]$ and $[-37/10000,0]$, respectively, polynomial coefficient bounds give $\ajj>0$ and $\frac72<y_n<\frac92$, so $H_n=\beta_{n,2,-2}$. On $\overline{B_n}$, coefficient bounds also give positivity of the $aa$ entry and determinant of the Hessian of $\Theta_1^2\Theta_2^2\beta_{n,2,-2}$, while Sturm counts give positivity of this mode on the three edges away from the BPS curve.

On the BPS edge, the mode vanishes only at $a_c(n,2)$ by \eqref{eq:alongcurve}. Its violating component is therefore confined to $B_n$, with closure meeting the BPS curve only there. Strict convexity makes this island convex and excludes any other BF-violating component of the level in $B_n$.

\phantomsection\label{app:profile}
Fix $I\Subset(0,1)$ and use $a_0\in I$, $j_n$, $a_n$, $c_0$, $\tauh_0$ and the bounded scaled variables $(\hat u,\hat v)$ of section~\ref{sec:highdents}. The estimates below are uniform in $a_0\in I$, but not as $a_0\to0,1$. Equations~\eqref{eq:mufact}, \eqref{eq:bpscoeffs} and \eqref{eq:Pderiv} express the coefficients in \eqref{eq:spinsquare} through analytic functions $Y_0,Y_1,d_0,d_1,d_2$ as
\begin{equation}
\begin{gathered}
y_n=QY_1+Y_0,\qquad \mu_n=\eta(d_2Q^2+d_1Q+d_0),\\
\ajj\big|_{\eta=0}=4c^2,\quad y_n\big|_{\eta=0}=\zeta_n,\quad d_2\big|_{\eta=0}=\tauh.
\end{gathered}
\end{equation}
Their derivatives are bounded on a fixed neighborhood of $I\times\{0\}$, where $\ajj,Y_1,d_2$ have positive lower bounds. Since $|a_n-a_0|=O(n^{-1})$ and $\zeta_n'=2(2n+1)/(1-a)^2$, the scaling gives
\begin{equation}
\mu_n=\tauh_0\hat v+O(n^{-1}),\qquad \frac{\ajj}{4}=c_0^2+O(n^{-1}),\qquad y_n-2j_n=\frac{4\hat u}{(1-a_0)^2}+O(n^{-1}).
\end{equation}
For large $n$, $y_n>0$, and periodicity and Lipschitz continuity of $\dist(\cdot,\mathbb Z)^2$ in \eqref{eq:levelmin} give \eqref{eq:profile}.

To locate the finite-level boundary, take $V>\max_Ic(a)^2/[4\tauh(a)]$, exceeding $|\hat v_\infty|$ in \eqref{eq:vinf}. For competing spins, $|2j-y_n|\le1$ and $\partial_{\hat v}\beta_{n,j,-j}=\tauh_0+O(n^{-1})>0$; minimization over spins preserves this lower slope. Together with $H_n(\hat u,-V)<0<H_n(\hat u,V)$ for large $n$, this gives a continuous zero-set graph $\hat v_n(\hat u)$ in $|\hat v|\le V$. On compact $\hat u$-intervals, $\sup|\hat v_n-\hat v_\infty|=O(n^{-1})$, and level $n$ satisfies the bound exactly when $\hat v\ge\hat v_n$. Equation~\eqref{eq:alongcurve} gives $\hat v_n\le0$, with equality exactly at saturation points.

Outside the scaled window, for fixed small $\delta>0$ and with $V$ enlarged if necessary, the coefficient bounds above give $\mu_n\le-\kappa|\eta|n^2$ ($\kappa>0$) on $-\delta\le\eta\le-V/n^2$, so $H_n\le\mu_n+\ajj/16<0$ there by \eqref{eq:roundbound}. Hence, for large $n$ and uniformly on $I$, level $n$ satisfies the bound just above the BPS curve only in its dents.

For the deepest point of a dent, put $\hat x=4\hat u/(1-a_0)^2$; for competing spins,
\begin{equation}
\begin{aligned}
\partial_{\hat u}\beta_{n,j,-j}&=\frac{8c_0^2}{(1-a_0)^2}[\hat x-2(j-j_n)]+O(n^{-1}),\\
\partial_{\hat u}^2\beta_{n,j,-j}&=\frac{32c_0^2}{(1-a_0)^4}+O(n^{-1}),\qquad \partial_{\hat u}\partial_{\hat v}\beta_{n,j,-j}=O(n^{-1}).
\end{aligned}
\end{equation}
The positive transverse derivative makes the graph uniformly Lipschitz. Along it, $dy_n/d\hat u=4/(1-a_0)^2+O(n^{-1})>0$, so exactly one tie, at $y_n=2j+\frac12$, separates adjacent saturation points. On each side, $H_n$ is one of the two tying modes. Its $\partial_{\hat u}\beta$ increases strictly and vanishes at its saturation point, so the two branches descend to the tie with nonzero opposite slopes. This kink is the unique deepest point. There $\dist(\hat x,\mathbb Z)=\frac12+O(n^{-1})$, and \eqref{eq:profile} gives $\eta_{\rm dent}=-c_0^2/(4\tauh_0n^2)+O(n^{-3})$.

\section{BF bounds for the complete \texorpdfstring{$t^+$}{t+} tower}\label{app:D}

We now establish the whole-tower bounds of sections~\ref{sec:wholetower} and~\ref{sec:large}. The final subsection also supplies the anti-aligned results used in sections~\ref{sec:branch} and~\ref{sec:kmin}. Elsewhere the background lies in $\Dom$, so the minimization in appendix~\ref{app:kmin} selects $k=-j$. We use the quadratic form \eqref{eq:form}, its coefficients \eqref{eq:coeffs} and determinant \eqref{eq:Delta}, together with the fixed-level minimum \eqref{eq:levelmin} and the completed square \eqref{eq:ellipse}.

The polynomial inequalities below use positivity of coefficients after a change of variables, Bernstein expansions on compact rectangles, and Sturm's theorem for univariate factors~\mbox{\cite{Farouki:2012,Garloff:1986,Basu:2006}}. In a Bernstein expansion, nonnegative coefficients imply nonnegativity on the rectangle; positivity on an open face follows when its restricted expansion has a positive coefficient.\footnote{The accompanying archive \texttt{Liu\_Miao\_Tang\_ancillary.zip} contains the polynomial data and exact checks used here and in appendix~\ref{app:islands}, together with the polynomials $c_i(a)$ in \eqref{eq:NPsi}.}

\subsection{Finite truncations above the BPS curve}\label{app:above}

Fix a rotation $a_0$ at which no mode saturates on the BPS curve, and fix $N,J\ge0$. Equations~\eqref{eq:sos} and \eqref{eq:alongcurve} make every mode positive at $(a_0,0)$. Since $\ajj\to4c(a_0)^2>0$ by \eqref{eq:bpscoeffs}, with $c(a)$ defined in \eqref{eq:zeta}, only finitely many spins compete at each level nearby. Each $H_n$ with $n\le N$ is therefore continuous and positive in a neighborhood of $(a_0,0)$.

For the modes with $j\le J$, analyticity in $\eta$ gives $\kappa_0>0$ such that $|\beta_{n,j,-j}(a_0,\eta)-\beta_{n,j,-j}(a_0,0)|\le\kappa_0|\eta|(1+Q^2)$ for small $|\eta|$. The BPS vertex $\zeta_n$ is linear in $n$, so there are $\gamma_0>0$ and $n_1$ with $\beta_{n,j,-j}(a_0,0)=c(a_0)^2(\zeta_n-2j)^2\ge\gamma_0Q^2$ for $n\ge n_1$. These modes satisfy the bound when $|\eta|$ is sufficiently small and below $\gamma_0/(2\kappa_0)$. The remaining finitely many modes with $j\le J$, $n<n_1$ remain positive by continuity. Combining these neighborhoods with those for $n\le N$ gives the $\delta$ of section~\ref{sec:above} for $-\delta<\eta<0$.

At $a=\frac15$ put $s=7/5+\eta$, $d_5=225(2+\eta)^2(4+5\eta)^2(12+25\eta)^2>0$, and define $p_0$, $p_1$, $p_2$ by
\begin{equation}
\label{eq:firstlevelpolys}
\begin{gathered}
\beta_{n,n/2,-n/2}=\frac{\eta p_2n^2+\eta p_1n+p_0}{d_5},\qquad p_0=d_5(4\aQQ+2\aQ+1),\\
\eta p_2=d_5(\aQQ+\aQj+\tfrac14\ajj),\qquad \eta p_1=d_5(4\aQQ+2\aQj+\aQ+\tfrac12\aj).
\end{gathered}
\end{equation}
On $-1/100\le\eta<0$, Sturm counts give $p_2>0$, $p_1<0$, $p_0>0$, $\ajj>0$ and $\mu_n>0$ for $n\le2$. Hence levels $n\le2$ satisfy the bound, and the numerator has the unique positive root
\begin{equation}
\label{eq:firstlevelroot}
n_+(\eta)=\frac{-\eta p_1-\sqrt{\eta^2p_1^2-4\eta p_2p_0}}{2\eta p_2}.
\end{equation}
Sturm counts also give $2<n_+<1/(2|\eta|)-1$. Since $y_n$ is affine in $n$, they establish $|y_n-n|<\frac12$ for $3\le n\le1/(2|\eta|)$. Thus $2j=n$ minimizes every level from $n=3$ through the first violating level. The numerator is negative exactly for $n>n_+$, giving
\begin{equation}
\label{eq:firstlevel}
n_{\rm v}(\eta)=\lfloor n_+(\eta)\rfloor+1,\qquad 2j=n_{\rm v},\qquad k=-j.
\end{equation}
As $p_2/d_5$, $p_1/d_5$, $p_0/d_5$ equal $1/3$, $-31/15$, $1/25$ at $\eta=0$,
\begin{equation}
\label{eq:firstlevelasymptotic}
n_+(\eta)=\frac{\sqrt3}{5\sqrt{|\eta|}}+\frac{31}{10}+O\big(\sqrt{|\eta|}\big).
\end{equation}
Since $n_{\rm v}=n_++O(1)$, the leading term gives \eqref{eq:firstlevelmainasymptotic}. A truncation $n\le N$ detects violation at $(\frac15,\eta)$ exactly when $n_{\rm v}(\eta)\le N$, leaving the passing layer of width $\sim3/(25N^2)$ given by \eqref{eq:firstlevelasymptotic}.

\subsection{The charge--spin lattice criterion and lower bounds below the BPS curve}\label{app:below}

For $\eta>0$, we have $\ajj>0$ by \eqref{eq:Njjeta} and $\Delta>0$ by \eqref{eq:Delta}. The minimum in \eqref{eq:ellipse} is
\begin{equation}
\label{eq:bmin}
\begin{aligned}
\bmin=-\frac{\eta}{(2a+\eta)^2(4a+\eta)\,\Theta_1}\,\big(&32a^2+192a^3+192a^4+32a\eta+272a^2\eta+384a^3\eta\\
&+8\eta^2+124a\eta^2+278a^2\eta^2+18\eta^3+84a\eta^3+9\eta^4\big),
\end{aligned}
\end{equation}
whose bracket is $N_\beta$, and
\begin{equation}
\label{eq:Qadmissible}
\begin{aligned}
\Qmin-2={}&\frac{1}{(2a+\eta)^2(2+2a+\eta)(4a+\eta)}\,\big[16a^2(1+a)(1+2a)+4a(4+23a+25a^2)\,\eta\\
&+2(2+25a+48a^2)\,\eta^2+2(4+17a)\,\eta^3+4\eta^4\big]>0 .
\end{aligned}
\end{equation}
If $\jmin\ge0$, the minimum over real $Q\ge2$, $j\ge0$ is $\bmin<0$. As $\jmin=J_m/[2(1-a^2)(2a+\eta)^2(4a+\eta)]$ with $J_m=16a^3(1+8a+7a^2)+12a^2(1+16a+19a^2)\eta+4a^2(23+43a)\eta^2+(54a^2+14a-1)\eta^3+6a\eta^4$, $\jmin<0$ requires $54a^2+14a<1$, that is, $a<(\sqrt{103}-7)/54$.

By convexity, the minimum over $j\ge0$ then lies on the singlet line. It occurs at $Q_0^*=-\aQ/(2\aQQ)=\Qmin+(\aQj/\aQQ)\jmin>\Qmin$ and equals
\begin{equation*}
1-\frac{\aQ^2}{4\aQQ}=\frac{\eta\,(4a+\eta)}{\NQQ}\Big[-3(1-2a^2)\,w^2+4a^2\,w-4a^4\Big],\qquad w=s-1=2a+\eta ,
\end{equation*}
which is negative: $a<(\sqrt{103}-7)/54<\frac1{14}$ gives $a^2<\frac13$, where the bracket has a negative leading coefficient and discriminant $16a^4(6a^2-2)<0$.

By \eqref{eq:ellipse}, modes with $Q\notin\mathcal Q_-$ of \eqref{eq:finitecharges} have $\beta\ge0$, and at $Q\in\mathcal Q_-$ the minimum, at the point of $\frac12\Zp$ nearest $\jst(Q)$, is $\mathcal B(Q)$, proving \eqref{eq:finitecriterion}. For the full minimum of either sign, if $\bar\beta$ is the value of an admissible trial mode, modes with $\beta\le\bar\beta$ have charges in the closed interval
\begin{equation}
\label{eq:fullminimuminterval}
(Q-\Qmin)^2\le\frac{\ajj}{\Delta}(\bar\beta-\bmin),\qquad Q\in\mathbb Z_{\ge2}.
\end{equation}
Minimizing over spins at these charges gives the full minimum.

The bound $\beta>-3$ of section~\ref{sec:ellipse} holds on $\eta=0$ by \eqref{eq:sos}. For $\eta>0$ it follows from the identities for the singlet-line minimum and for $\bmin$, each plus $3$,
\begin{equation}
\label{eq:floorsinglet}
\begin{aligned}
1-\frac{\aQ^2}{4\aQQ}+3=\frac{8a^2}{\NQQ}\,\big(&24a^2+48a^3+24a^4+24a\eta+88a^2\eta+64a^3\eta\\
&+6\eta^2+48a\eta^2+64a^2\eta^2+8\eta^3+24a\eta^3+3\eta^4\big)>0,
\end{aligned}
\end{equation}
and
\begin{equation}
\label{eq:floormin}
\bmin+3=\frac{8(1-a^2)(2a+\eta)^2(4a+\eta)\,\jmin+\Xi}{(2a+\eta)^2(4a+\eta)\,\Theta_1},
\end{equation}
where $\Xi$ is the polynomial with positive coefficients
\begin{equation}
\label{eq:Xi}
\Xi=32a^3+64a^4+32a^5+40a^2\eta+48a^3\eta+72a^4\eta+16a\eta^2+8a^2\eta^2+32a^3\eta^2+2\eta^3+4a^2\eta^3 .
\end{equation}
If $\jmin\ge0$, then $\beta\ge\bmin>-3$; if $\jmin<0$, the minimum over $j\ge0$ lies on $j=0$, where \eqref{eq:floorsinglet} applies.

For \eqref{eq:linearstrength} and its $g(a)$, let $D_\beta=(2a+\eta)^2(4a+\eta)\Theta_1$ be the positive denominator of \eqref{eq:bminform}; then
\begin{equation*}
\bmin+g(a)\eta =\frac{\eta^2 N_g(a,\eta)}{a(1+a)(1+5a)D_\beta},
\end{equation*}
where $N_g$ has positive coefficients, proving \eqref{eq:linearstrength}. As $N_\beta/D_\beta\to g(a)$ for $\eta\to0^+$, uniformly on compact subintervals of $(0,1)$, the coefficient is sharp for the continuous minimum.

\subsection{One-sided neighborhoods and excluded measure just below the BPS curve}\label{app:reach}

The local structure of $\Cset$ and the excluded set $V_\eta$ follows from a uniform positive bound away from $\Cset$ and violating modes at its points. For every compact $I\subset(0,1)\setminus\Cset$, there are $\gamma_I,\eta_I>0$ such that $\beta_{n,j,k}\ge\gamma_I\,\eta$ for all modes, $a\in I$ and $0<\eta\le\eta_I$. Conversely, every $\bar a\in\Cset$ is the saturation point of a mode with $P<0$. The only zero of $P$ is $(n,j)=(4,13)$, but its rotation $a_c(4,13)=\frac35=a_c(0,1)$ also has $P(0,1)=-4$ (appendix~\ref{app:sos}). Equation~\eqref{eq:tangexp} at $u=0$ then supplies a violating mode at each $(\bar a,\eta)$ for sufficiently small $\eta>0$.

By appendix~\ref{app:sos}, $P(n,j)\le0$ requires $n\le3$, $j\le\frac32$, or $n\ge4$ with $2\le j\le\frac12T_n\le13$, where $T_n$ is given in \eqref{eq:Tn}. For each $j\ge2$, only finitely many levels qualify. Since $a_c(n,j)$ increases to $1$ with $j$ and decreases to $0$ with $n$, $\Cset$ is a rational set, finite on compact subsets of $(0,1)$, with accumulation only at $a=0$ through $j\le\frac32$ and at $a=1$ through $n\le3$. The low-spin and low-level signs in section~\ref{sec:leaving} place $1/(4n+3)$ and $(1+2j)/(3+2j)$ in $\Cset$. They also supply $1/(4m+4)=a_c(2m+1,\frac12)$ and $1/(4m+6)=a_c(2m+2,\frac12)$ between each consecutive pair of singlet saturation points $1/(4m+7)$ and $1/(4m+3)$, with $m\in\Zp$.

For the uniform lower bound, fix $\eta_1>0$ and write $\GG$ in \eqref{eq:levelsquare} as $G_2j^2+G_1j+G_0$. On $I\times[0,\eta_1]$, \eqref{eq:G2} gives $G_2\ge G_{\min}>0$, while $|G_1|,|G_0|\le G_{\max}$. Consequently $\GG\ge1$, and hence $\beta\ge\eta$, for $j\ge j_*=\big(G_{\max}+\sqrt{G_{\max}^2+4G_{\min}(G_{\max}+1)}\big)/(2G_{\min})$.

Only finitely many half-integer spins $j<j_*$ remain. For each, the minimum over levels, $H^{(j)}=\aQQ\,\dist(v_j,\Zp)^2+\eta\,\GG(j)$, is continuous. Near $a_1\in I$ with $\GG(j;a_1,0)>0$, one has $H^{(j)}\ge\frac12\GG(j;a_1,0)\,\eta$. Otherwise $v_j(a_1,0)\notin\Zp$: a nonnegative integer value would make $a_1$ a saturation point with $P\le0$ by \eqref{eq:P}, contrary to $a_1\notin\Cset$. In that case $e_0=H^{(j)}(a_1,0)>0$, so nearby $H^{(j)}\ge\big(e_0/(2\eta_1)\big)\eta$. A finite subcover of $I\times\{0\}$ gives common constants $\gamma_I$ and $\eta_I$.

Since $\Cset$ is closed in $(0,1)$, the uniform bound gives a one-sided neighborhood in $S$ at every $a_0\notin\Cset$; the saturating modes exclude such a neighborhood at points of $\Cset$. Define $A_m=\{a:(a,\eta)\in S\ \text{for}\ 0<\eta\le1/m\}$ for $m\ge1$. These sets are closed in $(0,1)$ because $S$ is closed in $\Dom$, increase with $m$, and have union $(0,1)\setminus\Cset$, of full measure. For $0<\eta\le1/m$, the passing rotations include $A_m$. Their measure therefore tends to one as $\eta\to0^+$.

For the excluded measure, we define $w(\bar a)$, for $\bar a\in\Cset$, by
\begin{equation}
\label{eq:wmeasure}
w(\bar a)^2=\max_{\substack{a_c(n,j)=\bar a\\P(n,j)<0}} \frac{4|P(n,j)|(n+3j+2)}{(n+j+1)(4n+2j+3)^3},
\end{equation}
and we show that, in a fixed small neighborhood of $\bar a$ and for small $\eta>0$, the violating rotations form one interval with endpoints
\begin{equation}
\label{eq:measureendpoints}
a_\pm(\eta)=\bar a\pm w(\bar a)\sqrt\eta+O(\eta),\qquad\eta\to0^+,
\end{equation}
so that, if $I\Subset(0,1)$ has endpoints outside $\Cset$,
\begin{equation}
\label{eq:measurelocal}
|V_\eta\cap I|=2\sqrt\eta\sum_{\bar a\in\Cset\cap I}w(\bar a)+O_I(\eta),
\end{equation}
and near $a=0$ the singlets give
\begin{equation}
\label{eq:measurelower}
|V_\eta|\ge\tfrac12\eta^{1/3},\qquad 0<\eta\le1/125.
\end{equation}
Near $\bar a$, only its saturating modes compete. This follows from coercivity in the level, continuity and the large-spin bound above, which also holds near $\bar a$ because it uses only \eqref{eq:G2}. Each competing mode has $\partial_a^2\beta=2K_{n,j}+O(|a-\bar a|+\eta)>0$, with $K_{n,j}$ and $\Sigma_{n,j}$ given in \eqref{eq:tangexp}--\eqref{eq:P}. At small fixed $\eta$, its negative set is empty for $P>0$. For $P<0$, it is an interval around $\bar a$ with endpoints $\bar a\pm\sqrt{|\Sigma_{n,j}|/K_{n,j}}\sqrt\eta+O(\eta)$. At $\bar a=\frac35$, the saddle \eqref{eq:degenerate} contributes only width $O(\eta)$, inside the interval of $(0,1,-1)$. Taking the union gives \eqref{eq:measureendpoints}. Summing over the finite set $\Cset\cap I$, and applying the uniform lower bound elsewhere on $I$, gives \eqref{eq:measurelocal}.

For \eqref{eq:measurelower}, the singlet at an integer nearest $Q_0^*$ has $\beta\le\Psi_0\equiv1-\aQ^2/(4\aQQ)+\aQQ/4$ (appendix~\ref{app:below}). After the substitution $a=\tau_1\tau_2/10$, $\eta=\tau_2^3/125$, which maps $(0,1]^2$ onto $0<\eta\le1/125$, $0<a\le(\eta/8)^{1/3}=\frac12\eta^{1/3}$, Bernstein coefficients give $\Psi_0<0$ and $2Q_0^*-3>0$, so this admissible singlet violates the bound.

\subsection{The outer strip and the strength of violation within it}\label{app:width}

The rounding estimate in section~\ref{sec:width} requires admissible quantum numbers and a uniform control of the sign of $\Psi$. Its denominator is positive by \eqref{eq:etaforms} and \eqref{eq:Njjeta}; we first address admissibility.

Equation~\eqref{eq:Qadmissible} ensures $Q_\circ\ge2$. Since $|Q_\circ-\Qmin|\le\frac12$ and $\rho\le0$, one has $\jst(Q_\circ)\ge\jmin+\rho/2$, so the rounded spin is nonnegative if $j_{\rm ad}\equiv\jmin+\rho/2+\frac14>0$. Its sign is that of an integer polynomial. For $a\ge\frac1{29}$, the negative monomials $-2\eta^5-6\eta^6-3\eta^7$ are dominated by $60a\eta^5+174a\eta^6+143a\eta^7$; for $\eta<12a$, positivity follows after substituting $\eta=12az/(1+z)$.

In the remaining corner, use the singlet at an integer nearest $Q_0^*$, with $\beta\le\Psi_0$ as in appendix~\ref{app:reach}. Substituting $a=z_1/(29(1+z_1))$ and $\eta=a(12+z_2)$, with $z_1>0$ and $z_2\ge0$, gives $\Psi_0<0$ and $2Q_0^*-3>0$ by coefficient signs. This singlet is therefore admissible and violates the bound.

By a Sturm count, $N_\Psi(a,9.37868\times10^{-3})$ has no root on $[0,1]$ and is negative at $a=1/2$. As $N_\Psi$ decreases in $\eta$, this gives $\eta_*(a)<9.37868\times10^{-3}$ and hence \eqref{eq:stripmaximum}. By compactness the negativity persists at a slightly smaller $\eta$, so the supremum of $\eta_*$ is strictly below $9.37868\times10^{-3}$, and $N_\Psi(4/35,9.37867\times10^{-3})>0$ gives
\begin{equation*}
9.37867\times10^{-3}<\sup_{0<a<1}\eta_*(a)<9.37868\times10^{-3}.
\end{equation*}
At $\eta=4a/15$, $N_\Psi$ is a negative prefactor times a polynomial with positive coefficients, so $\eta_*(a)<4a/15<12a$ and the rounded mode is admissible on the curve $\eta=\eta_*(a)$; there $\beta<\Psi=0$ whenever $\Qmin\notin\mathbb Z+1/2$, the charge rounding cost being below $\Delta/(4\ajj)$.

For \eqref{eq:stripfloor}, let $0<a\le\frac1{25}$ and $t=\eta/a$, so $0<t<4/15$ in the strip; with $D_\beta$ of appendix~\ref{app:below},
\begin{equation}
\label{eq:strengthpolynomials}
\begin{aligned}
D_\beta-4\eta N_\beta &=a^3\big[2(t+2)^2(4-15t)+a h_1(t)+a^2h_2(t)\big],\\
h_1(t)&=192-432t-872t^2-436t^3-66t^4,\\
h_2(t)&=160-376t-1168t^2-946t^3-300t^4-33t^5.
\end{aligned}
\end{equation}
Both $h_1$ and $h_1+h_2/25$ decrease for $t\ge0$ and are positive at $t=4/15$, so $h_1+ah_2$, affine in $a$, is positive for $0<a\le\frac1{25}$, $0\le t\le4/15$; hence $D_\beta-4\eta N_\beta>0$, which by \eqref{eq:bminform} means $\bmin>-\frac14$. For $a\ge\frac1{25}$, $g$ decreases and, by \eqref{eq:stripmaximum},
\begin{equation*}
-g(a)\eta>-0.0094\,g(1/25) =-\frac{36707}{156000}>-\frac14.
\end{equation*}
With \eqref{eq:linearstrength} this gives \eqref{eq:stripfloor}. For sharpness of $\frac14$ for the continuous minimum, expansion at $\eta=at$ as $a\to0$ gives
\begin{equation*}
a^{-7}N_\Psi(a,at)=-8(t+2)^6(15t-4)+O(a), \qquad \bmin(a,at)\longrightarrow-\frac{4t}{4+t},
\end{equation*}
uniformly for $0\le t\le4/15$; as $0<\eta_*(a)/a<4/15$, the first limit gives $\eta_*(a)/a\to4/15$ and the second $\bmin(a,\eta_*(a))\to-\frac14$.

\subsection{Uniform positivity at high level and spin}\label{app:highlevels}

For real $n\ge20$, $j\ge2$, write $Q=n+2\ge22$ and $N(Q,j)=\Theta_1^2\Theta_2^2\,\beta_{Q-2,j,-j}$. The auxiliary polynomial
\begin{equation}
\label{eq:mdef}
\mathcal M=a(1+s)(3s^2-2s-1-4a^2)\,U
\end{equation}
has only positive coefficients in $(a,\eta)$, as do $\Njj$, $\NQQ$, $\Theta_1$ and $\Theta_2$. The same holds for $(s-1)^2(s+1)(s-1+2a)$, the leading coefficient of the quadratic $\varpi(Q)$ in \eqref{eq:mufact}. By \eqref{eq:mufact} and $\Theta_1^2\Theta_2^2\ajj=8(1-a^2)^2\Njj$,
\begin{equation}
\label{eq:exclsquare}
\beta_{Q-2,j,-j}=\ajj\Big(j-\frac{y_n}2\Big)^2+\eta\,\frac{(1+s)\,\varpi(Q)}{\Njj \Theta_1},
\end{equation}
with $\ajj>0$ on the domain under consideration. The real-spin vertex lies in $j\ge2$ exactly when $y_n\ge4$. Since $-\aQj=4(1-a^2)\mathcal M/(\Theta_1^2\Theta_2^2)$ by \eqref{eq:coeffs}, this is equivalent to $Q\ge Q_v\equiv\mathcal V/\mathcal M$, with $\mathcal V=\Theta_1\Theta_2^2+4(1-a^2)\Njj$. At the boundary $j=2$, write $N(Q,2)=\mathcal E_2(Q-22)^2+\mathcal E_1(Q-22)+\mathcal E_0$, where $\mathcal E_2=(1+s)^2\NQQ$, $\mathcal E_1=8(1+s)\mathcal E_{1r}$, $\mathcal E_0=\Theta_1^2\Theta_2^2\beta_{20,2,-2}$ and
\begin{equation}
\label{eq:excldisc}
4\mathcal E_0\mathcal E_2-\mathcal E_1^2=4\eta\,(1+s)^2\Theta_2^2\Theta_1\,F_{\rm e} ,
\end{equation}
which defines $F_{\rm e}$.

At fixed $Q\ge22$, minimization over $j\ge2$ gives two alternatives. If $y_n\ge4$, the vertex is admissible and \eqref{eq:exclsquare} gives $\beta\ge\eta(1+s)\varpi(Q)/(\Njj\Theta_1)$. If $y_n<4$, the minimum occurs at $j=2$, so $\beta\ge N(Q,2)/(\Theta_1^2\Theta_2^2)$. Both $\varpi$ and $N(Q,2)$ are convex in $Q$; it remains to make these lower bounds positive on their respective charge ranges. On a covering of the $(a,\eta)$ domain $(0,1)\times(0,\infty)$, it therefore suffices to establish either
\begin{equation}
\label{eq:exclI}
F_{\rm e}>0,\qquad \Pi_v\equiv\frac{\mathcal M^2\varpi(Q_v)}{(1+s)\Njj}>0,\qquad \Pi_d\equiv\frac{\mathcal M\,\varpi'(Q_v)}{(1+s)\Njj}\ge0,
\end{equation}
so that $N(Q,2)>0$ for all real $Q$ and $\varpi>0$ on $[Q_v,\infty)$, or
\begin{equation}
\label{eq:exclII}
\mathcal E_0>0,\qquad \mathcal E_{1r}\ge0,\qquad \varpi(22)>0,\qquad \varpi'(22)\ge0,
\end{equation}
so that $N(Q,2)\ge\mathcal E_0$ and $\varpi(Q)\ge\varpi(22)$ for $Q\ge22$.

The integer polynomials in \eqref{eq:exclI} have nonnegative Bernstein coefficients on $R_1=[0,15/256]\times[0,1/2]$; those in \eqref{eq:exclII} do so on $R_2=[15/256,1]\times[0,1/2]$ and $R_3=[0,1]\times[1/2,\infty)$. For $R_3$, use $v\in[0,2]$ and replace each polynomial $f$ by $v^{\deg_\eta f}f(a,1/v)$. Zeros occur only in the $R_1$ arrays of $F_{\rm e}$, $\Pi_v$ and $\Pi_d$, which vanish at the unphysical corner $a=\eta=0$. The Bernstein positivity property stated above gives the required strict signs throughout the physical domain.

\subsection{Large-radius asymptotics and the level-four minimum}\label{app:large}

At fixed $0<a<1$, expansion of \eqref{eq:coeffs} and \eqref{eq:bmin} in $1/s$ gives \eqref{eq:largeS}. The rounded spin satisfies $|j_\circ-\jst(6)|\le\frac14$, and $6-\Qmin=6a/s+O(s^{-2})$, so \eqref{eq:ellipse} gives \eqref{eq:levelfour}. At $Q=6$, the minimum is attained at a half-integer $j$ nearest $\jst(6)\propto r_+$ and equals $\bmin+O(s^{-2})$. Using $s=\sqrt2\,r_+\big(1+O(r_+^{-2})\big)$ gives the level-four value in \eqref{eq:largemin} as $r_+\to\infty$.

To exclude other levels, \eqref{eq:ellipse} gives $\beta\ge\bmin+(\Delta/\ajj)(Q-\Qmin)^2$. For $Q\ne6$, the separation $|Q-\Qmin|\ge1-O(1/s)$ implies $\beta\ge-3+\frac19+O(1/s)$, above the level-four minimum. For the tail $Q\ge12$, $|Q-\Qmin|\ge6-O(1/s)$ gives $\beta\ge1-o(1)$ uniformly in $Q$. The level-four mode therefore attains the full minimum for sufficiently large radius.

For \eqref{eq:scalinglimit}, fix $Q$ and take $j=\varsigma s+O(1)$, $k=-j$. By \eqref{eq:nhconstants}, $\ell_2^2=B^2\to\frac1{12}$, $F^2/A^2\to\frac23$, $E=O(s^{-2})$, $s^2/C^2\to2(1-a^2)$, $s^2/D^2\to2(1-a^2)^2$ and $k+QG=\big[\frac{aQ}{2(1-a^2)}-\varsigma\big]s+O(1)$. Hence, in $\beta$ of \eqref{eq:beta} with \eqref{eq:m2q2}, the fiber term $16\ell_2^2(k+QG)^2/D^2$ tends to the last term of \eqref{eq:scalinglimit}, the terms with $j(j+1)-k^2$ or $E$ vanish, and the rest tends to $1+\frac13M^2-\frac29Q^2=\frac19\big[(Q-6)^2-27\big]$, the limit $r_+\to\infty$ of \eqref{eq:RNj} at fixed $j$. The minimum of \eqref{eq:scalinglimit} over $\varsigma\ge0$, at $\varsigma=aQ/[2(1-a^2)]$, is $H_n^\infty=(n^2-8n-11)/9$, negative exactly for $n\le9$; as $\beta_{n,j,-j}$ is quadratic in $j$, with $\varsigma^2$ coefficient tending to $\frac83(1-a^2)^2>0$, the level minimum $H_n$ tends to $H_n^\infty$.

For singlets, set $\varsigma=0$. The limit $\frac19[(1+6a^2)Q^2-12Q+9]$ increases with $Q\ge2$ when $a^2\ge\frac{11}{24}$ and equals $(24a^2-11)/9$ at $Q=2$. Together with the uniform tail bound, this proves the criterion of section~\ref{sec:largeradius} for $a^2\ne\frac{11}{24}$. At equality, the next term gives $\beta_{0,0,0}=(1320s-1511)/(1296s^2)+O(s^{-3})>0$.

\subsection{BF violation for both anti-aligned charge families}\label{app:antialigned}

The two families with $q_C<0$ are the first charge root beyond $q_C=0$ and the second root. By appendix~\ref{app:horexp}, \eqref{eq:betaclosed} holds on the first charge root for all $r_+>0$ and, at $s\to-s$, on the second, where $\Omega_H$ and $\Phi_H$ are likewise given by \eqref{eq:OmPhi} at $s\to-s$ (section~\ref{sec:branch}).

Neither family is supersymmetric, even in the type IIB uplift. With the CCLP charges, the four eigenvalues of the Bogomol'nyi matrix of the minimal theory are positive multiples of $m_C-(1\pm2a)q_C=(m_C+q_C)-2(1\pm a)q_C$ and of $m_C+q_C$~\mbox{\cite{Cvetic:2005zi}}. Writing $\bar s=\sqrt{1+2a^2+2r_+^2}>1$ for the positive square root on the second family, whose expressions are those of the first root at $s=-\bar s$, one has, on the first and second roots respectively,
\begin{equation}
m_C+q_C=\tfrac18(1+s)^2\big[(s-1)^2+4r_+^2\big],\qquad m_C+q_C=\tfrac18(\bar s-1)^2\big[(\bar s+1)^2+4r_+^2\big].
\end{equation}
So all four are positive for $q_C<0$. In the maximal theory, the three equal charges enter the sixteen eigenvalues separately~\mbox{\cite{Cvetic:2005zi}}. The twelve remaining eigenvalues are positive multiples of
\begin{equation}
\label{eq:maximaleigen}
(3\pm a)(m_C+q_C)-2(1-a^2)\,q_C,\qquad (3+a^2)(m_C+q_C)-4(1-a^2)\,q_C,
\end{equation}
hence also positive.

On the first family, set $w=s-1$. The conditions $q_C<0$ and $r_+>0$ give $2w^2+2w<4a^2<2w^2+4w$. The first inequality implies $s^2-s<2a^2<(1+2a)^2-(1+2a)$, hence $\eta<0$ and $\Omega_H>1$. Also, $\Phi_H=U/\Theta_2<0<(s^2-1)/\Theta_2=\Phi_H+1$. The spin coefficient follows from the factorization over $\mathbb Q(\sqrt3)$,
\begin{equation}
\label{eq:Njjfact}
\begin{gathered}
\Njj=-\tfrac12\,f_-f_+,\\
f_+=(3+\sqrt3)\,w^2+(4+2\sqrt3)\,w-4a^2,\qquad f_-=4a^2-w\big((3-\sqrt3)\,w+4-2\sqrt3\big),
\end{gathered}
\end{equation}
and since $f_+$ decreases and $f_-$ increases with $a^2$, here $f_+>w\big((1+\sqrt3)w+2\sqrt3\big)>0$ and $f_->(\sqrt3-1)w(w+2)>0$, so $\ajj<0$, the coefficient of $j^2$ in $\beta_{n,j,\pm j}$ at every level. As $3s^2-2s-1-4a^2=(s-1)^2+4r_+^2>0$ and $U<0$, the right-hand side of \eqref{eq:kdiff} is negative for $jQ>0$, which with the concavity \eqref{eq:kcurv} gives $k=+j$.

On the second family, with $s=-\bar s$, $\Omega_H-1=-(1-a)(\bar s+1+2a)/(\bar s+1-2a^2)$ and $\Phi_H+1=-(\bar s^2-1)/(\bar s+1-2a^2)$, where $\bar s\pm1$, $\bar s+1\pm2a$, $\bar s+1-2a^2$, $\Theta_1=3\bar s^2-1-2a^2$, $\bar s^2+\bar s-2a^2$ and $3\bar s^2+2\bar s-1-4a^2$ are positive; hence $\Omega_H<1$ and $\Phi_H<-1$. Moreover, \eqref{eq:kdiff} is $-16a(1-a^2)jQ$ times a quotient of these factors, so $k=+j$ by \eqref{eq:kcurv}.

At $k=+j$, $\beta$ is quadratic in $(Q,j)$ with constant term $1$ and the coefficients \eqref{eq:coeffs} at $s=-\bar s$, except that $\aQj$ changes sign; henceforth, $\aQQ,\dots,\aj$, $\Delta$, $\bmin$, $\Qmin$, $\jmin$, $\jst$ and $\rho=\aQj/\ajj$ refer to this polynomial. Then $\aQj<0$ and
\begin{equation}
\label{eq:Deltasheet2}
\Delta=\frac{8(1-a)^2(1+a)^2(\bar s-1)^2(\bar s+1)^2(\bar s+1-2a)(\bar s+1+2a)}{(\bar s+1-2a^2)^2(3\bar s^2-1-2a^2)^3}>0 .
\end{equation}

On the second family, $0<a<1$ and $\bar s>\sqrt{1+2a^2}$ imply $\Njj>0$ at $s=-\bar s$, so $\ajj>0$ and \eqref{eq:ellipse} applies. Moreover, $\Qmin>\frac52$ ensures that an integer $Q^*$ nearest $\Qmin$ satisfies $Q^*\ge3$. If $\jst(Q^*)\ge-\frac14$, choose $j^*\in\frac12\Zp$ nearest $\jst(Q^*)$. Then $\beta_{Q^*-2,j^*,j^*}\le\Psi<0$, with \eqref{eq:Psi} evaluated using the second-family coefficients.

Otherwise $0<-\jst(Q^*)\le-\jmin-\rho/2$, requiring $\jmin+\rho/2+\frac14<0$. The singlet obeys $\beta_{Q^*-2,0,0}\le\Psi_s\equiv\bmin+\Delta/(4\ajj)+\ajj(\jmin+\rho/2)^2$. Since $\jmin+\rho/2+\frac14>0$ for $a\ge\frac3{10}$, this case is confined to $a<\frac3{10}$, where $\Psi_s<0$. Thus every black hole in this family has an admissible violating mode.

The sign conditions used to select an admissible rounded mode or a violating singlet reduce, after removing positive factors, to polynomial inequalities in $(a,\bar s)$. Expanding the resulting polynomials in $y=\bar s-1-\frac7{10}a^2$, Sturm counts establish positivity of the coefficients on the stated rotation intervals, including the common endpoint $a=\frac3{10}$. The expansion variable satisfies $y>0$, since $1+2a^2-\big(1+\frac7{10}a^2\big)^2=a^2(60-49a^2)/100>0$.

\bibliographystyle{JHEP}
\bibliography{references}
\end{document}